\documentclass[12pt]{article}
\usepackage{amssymb,amsfonts,latexsym}

\newcommand{\sgb}{\mbox{\scriptsize{\gb}}}
\newcommand{\diag}{{\textup{\scriptsize{diag}}}}
\newcommand{\lr}{{\mathcal{L}}}
\newcommand{\fr}{{\mathcal{F}}}

\renewcommand{\j}{{\mathcal{J}}}
\newcommand{\g}{{\mathcal{G}}}
\newcommand{\gh}{g}

\newcommand{\hl}{\lfloor\frac{\lambda}{2}\rfloor}

\renewcommand{\theequation}{\arabic{section}.\arabic{equation}}

\newcounter{saveeqn}
\newcounter{savealpheqn}

\newcommand{\alpheqn}{\setcounter{saveeqn}{\value{equation}}%
 \stepcounter{saveeqn}\setcounter{equation}{0}%
 \renewcommand{\theequation}{\mbox{\arabic{section}.\arabic{saveeqn}\alph{equation}}}
 \renewcommand{\)}{\end{equation}}}
\newcommand{\reseteqn}{\setcounter{equation}{\value{saveeqn}}%
	\renewcommand{\theequation}{\arabic{section}.\arabic{equation}}%
	\renewcommand{\)}{\end{equation}}}
\def\getletter#1{\renewcommand{\theequation}{\alph{equation}}%
		 \begin{eqnarray}%
                 \label{#1}%
		 \nonumber\end{eqnarray}\vspace{-.666in}%
                 \renewcommand{\theequation}{\mbox{\arabic{section}.\arabic{saveeqn}\alph{equation}}}}
\def\writeletter#1{\renewcommand{\theequation}{\alph{#1}}%
		 \begin{eqnarray}%
                 \label{#1}%
		 \nonumber\end{eqnarray}\vspace{-.9in}}%
\def\getlette2r#1{\newcounter{#1}%
		 \setcounter{#1}{\value{equation}}%
	         \providecommand{\writeletters}{\writeletters\writeletter{#1}}}
\def\agetletter#1{\renewcommand{\theequation}{\alph{equation}}%
		 \begin{eqnarray}%
                 \label{#1}%
		 \nonumber\end{eqnarray}\vspace{-.666in}%
                 \renewcommand{\theequation}{\mbox{\Alph{subsection}.\arabic{saveeqn}\alph{equation}}}}
\newcommand{\aalpheqn}{\setcounter{saveeqn}{\value{equation}}%
 \stepcounter{saveeqn}\setcounter{equation}{0}%
 \renewcommand{\theequation}{\mbox{\Alph{subsection}.\arabic{saveeqn}\alph{equation}}}
  \renewcommand{\)}{\end{equation}}}
\newcommand{\areseteqn}{\setcounter{equation}{\value{saveeqn}}%
 \renewcommand{\theequation}{\Alph{subsection}.\arabic{equation}}%
 \renewcommand{\)}{\end{equation}}%
 }

\renewcommand{\thefootnote}{\alph{footnote}}
\renewcommand{\(}{\begin{equation}}
\renewcommand{\)}{\end{equation}}
\newcommand{\ba}{\begin{eqnarray}}
\newcommand{\ea}{\end{eqnarray}}
\renewcommand{\l}{\lambda}

\renewcommand{\a}{\alpha}
\renewcommand{\b}{\beta}
\renewcommand{\r}{\rho}
\newcommand{\sa}{\mathop{\vtop{\ialign{##\crcr
  $\hfil\displaystyle{\longleftrightarrow}\hfil$\crcr\noalign{\kern-1pt\nointerlineskip}
  \hspace{.12in}$^\sigma$\hskip6pt\crcr\noalign{\kern3pt}}}}}
\newcommand{\sat}{\mathop{\vtop{\ialign{##\crcr
  $\hfil\displaystyle{\longrightarrow}\hfil$\crcr\noalign{\kern-1pt\nointerlineskip}
  \hspace{.12in}$^\sigma$\hskip6pt\crcr\noalign{\kern3pt}}}}}
\newcommand{\pa}{\mathop{\vtop{\ialign{##\crcr
  $\hfil\displaystyle{\oplus}\hfil$\crcr\noalign{\kern+1pt\nointerlineskip}
  \hspace{.08in}$^{\alpha=0}$\hskip6pt\crcr\noalign{\kern3pt}}}}}
\newcommand{\pan}{\mathop{\vtop{\ialign{##\crcr
  $\hfil\displaystyle{\oplus}\hfil$\crcr\noalign{\kern+2pt\nointerlineskip}
  \hspace{.03in} $^{\alpha}$\hskip6pt\crcr\noalign{\kern3pt}}}}}

\newcommand{\s}{\sigma}

\renewcommand{\sp}{,\hspace{.3in}}
\newcommand{\newsection}{\setcounter{equation}{0}\section}
\newcommand{\p}{^\prime}
\newcommand{\ws}{\omega (h_\s)}
\newcommand{\w}{\omega}
\newcommand{\ud}{(U^\dag)}
\newcommand{\usd}{U^\dag(\s)}
\renewcommand{\lor}{\frac{\l}{\r}}
\newcommand{\lors}{\frac{\l}{\r(\s)}}
\newcommand{\lra}{\leftrightarrow}
\newcommand{\mod}{{\textup{\scriptsize{ mod }}}}
\newcommand{\rmod}{{\textup{ mod }}}
\newcommand{\sr}{\sqrt{\r}}
\newcommand{\srs}{\sqrt{\r(\s)}}
\newcommand{\isr}{\frac{1}{\sr}}
\newcommand{\isrs}{\frac{1}{\srs}}
\newcommand{\reg}{O((z-w)^0)}
\newcommand{\jh}{\hat{J}}
\newcommand{\tp}{{2\pi i}}
\newcommand{\az}{\frac{A(\zl)}{\zl}}
\newcommand{\ad}{\frac{A(\dl)}{\zl}}
\newcommand{\taz}{$A(\zl)/\zl$}
\newcommand{\tad}{$A(\dl)/\zl$}
\newcommand{\bu}{$\bullet$}
\newcommand{\dngg}{\textup{dim}g}
\newcommand{\dg}{\textup{dim}\gb}
\newcommand{\arange}{a,b=1,...,\dg}
\newcommand{\appendixa}
 {\renewcommand{\theequation}{\Alph{subsection}.\arabic{equation}}
  \renewcommand{\thesubsection}
               {Appendix \Alph{subsection}.\setcounter{equation}{0}}
  \renewcommand{\alpheqn}{\aalpheqn}
  \renewcommand{\reseteqn}{\areseteqn}}
\newcommand{\appendices}{\appendix\appendixa}

\def\gb            {\mbox{$\mathfrak g$}}
\def\sz		   {\mbox{\scriptsize $\mathbb  Z$}}
\def\z		   {\mbox{$\mathbb  Z$}}
\def\zl		   {\mbox{$\mathbb  Z_\l$}}
\def\subsecz	   {\mbox{\large $\mathbb  Z$}}
\def\mz		   {\mbox{\Large $\mathbb  Z$}}

\def\d             {\mbox{$\mathbb D$}}
\def\dl		   {\mbox{$\mathbb D_\l$}}
\def\subsecd	   {\mbox{\large $\mathbb  D$}}

\mathchardef\endbar="375
\font\fivesans=cmss10 at 4.61pt
\font\sevensans=cmss10 at 6.81pt
\font\tensans=cmss10 at 12pt 
\newfam\sansfam
\textfont\sansfam=\tensans\scriptfont\sansfam=\sevensans\scriptscriptfont
\sansfam=\fivesans

\font\tensans=cmss10 at 14pt 
\newfam\sansfamb
\textfont\sansfamb=\tensans\scriptfont\sansfamb=\sevensans\scriptscriptfont
\sansfamb=\fivesans

\font\tensans=cmss10 at 17pt
\newfam\sansfamc
\textfont\sansfamc=\tensans\scriptfont\sansfamc=\sevensans\scriptscriptfont
\sansfamc=\fivesans

\font\tensans=cmss10 at 10pt
\def\contr#1#2{\mathop{\vtop{\ialign{##\crcr
  $\hfil\displaystyle{#2}\hfil$\crcr\noalign{\kern3pt\nointerlineskip}
  \hspace{.09in}\rule[0in]{.01in}{.1in}\rule[0in]{#1in}{.01in}\rule[0in]{.01in}{.1in}\hskip6pt\crcr\noalign{\kern3pt}}}}}
\def\contrb#1#2#3{\mathop{\vtop{\ialign{##\crcr
  $\hfil\displaystyle{#3}\hfil$\crcr\noalign{\kern3pt\nointerlineskip}
  \hspace{#1in}\rule[0in]{.01in}{.1in}\rule[0in]{#2in}{.01in}\rule[0in]{.01in}{.1in}\hskip6pt\crcr\noalign{\kern3pt}}}}}
\def\namegroup#1{\begin{eqnarray}\label{#1}\nonumber\end{eqnarray}\vspace{-.616in}}
\def\hsp#1{\hspace{#1in}}
\def\rf#1{\ref{ref#1}}
\def\comment#1{\hsp{.3}\textup{#1}}

\catcode`\@=11
\def\vereq#1#2{\lower3pt\vbox{\baselineskip1.5pt \lineskip1.5pt
\ialign{$\m@th#1\hfill##\hfil$\crcr#2\crcr\sim\crcr}}}
\catcode`\@=12

\begin{document}
\begin{titlepage}
\begin{center}
August 27, 1999           \hfill UCB-PTH-99/36   \\
                               \hfill LBNL-44181    \\
                                \hfill hep-th/9908187    \\

\vskip .25in
\def\thefootnote{\fnsymbol{footnote}}
{\large \bf New Duality Transformations in Orbifold Theory \\}
\vskip 0.5in

J. de Boer$^{1,2}$, J. Evslin$^3$, M. B. Halpern$^3$, and J. E. Wang$^3$\footnote{E-Mail: hllywd2@physics.berkeley.edu}

\vskip 0.2in

$^1${\em Spinoza Institute, University of Utrecht,\\
Leuvenlaan 4, 3584 CE Utrecht, The Netherlands}\\

\vskip 0.2in

$^2${\em Instituut-Lorentz for Theoretical Physics, University of Leiden,\\
PO Box 9506, NL-2300 RA, Leiden, The Netherlands}\\

\vskip .2in

$^3${\em Department of Physics, University of California and\\
     Theoretical Physics Group, Lawrence Berkeley National Laboratory\\
     Berkeley, California 94720, USA}
        
\end{center}

\vskip .3in

\vfill

\begin{abstract}
We find new duality transformations which allow us to construct the stress tensors of all the twisted sectors of any orbifold $A(H)/H$, where $A(H)$ is the set of all current-algebraic conformal field theories with a finite symmetry group $H\subset Aut(g)$.  The permutation orbifolds with $H=\z_\l$ and $H=S_3$ are worked out in full as illustrations but the general formalism includes both simple and semisimple $g$.  The motivation for this development is the recently-discovered orbifold Virasoro master equation, whose solutions are identified by the duality transformations as sectors of the permutation orbifolds \tad. 
\end{abstract}

\vfill

\end{titlepage}
\setcounter{footnote}{0}
\renewcommand{\thefootnote}{\alph{footnote}}

%
%
%
%
%
%

\pagebreak
\renewcommand{\thepage}{\arabic{page}}
\newsection{Introduction}
Recently, an orbifold Virasoro master equation (OVME) was found which summarizes the general Virasoro construction$^{\rf{5}}$
\(
\hat{T}(z)=\sum_{r=0}^{\l-1}\lr_r^{ab}:\hat{J}_a^{(r)}(z) \hat{J}_b^{(-r)}(z):
 \label {introOVMET}
\)
on the orbifold affine algebra$^{\rf{9}}$ $\gb_\l$ at order $\l$.  The OVME at $\l=1$ is the Virasoro master equation (VME) which summarizes the general affine-Virasoro construction$^{\rf{1}-\rf{3}}$
\(
T(z)=L^{ab}:J_a(z) J_b(z): \label{introsimpleT}
\)
on affine Lie algebra$^{\rf{10}-\rf{12}}$.  The coefficients $\lr$ and $L$ in (\ref{introOVMET}) and (\ref{introsimpleT}) are called inverse inertia tensors.  It was clear for $\l\geq 2$ that the OVME collects a large class of stress tensors $\hat{T}$ of twisted sectors of orbifolds, but, except in special cases, these orbifolds were not identified in Ref.~\rf{5}.

In the present paper we find sets of \textit{new duality transformations} which provide relations among the sectors of the general current-algebraic orbifold.  In particular, the new duality transformations solve the identification puzzle of the OVME and allow the systematic construction of the stress tensors of all the sectors of any current-algebraic orbifold.  

The new duality transformations can be viewed as a synthesis of standard orbifold theory (see Refs.~\rf{28}-\rf{17}, \rf{9}, \rf{41} and \rf{42}) and the theory of the $H$-invariant CFT's (see Refs.~\rf{4}, \rf{47}, and \rf{3}) in the VME: The set $A(H)$ of $H$-invariant CFT's on $g$ collects the most general affine-Virasoro construction $T(L)$ with a finite symmetry group $H\subset Aut(g)$, where the Lie algebra $g$ can be simple or semisimple.  Each stress tensor $T(L)$ in $A(H)$ can also serve as the stress tensor of the untwisted sector of an orbifold $A(H)/H$.  For the stress tensors $\hat{T}_\s(\lr)$ of the twisted sectors $\s$ of $A(H)/H$, the new duality transformations 
\(
L \rightarrow \lr(\s;L)
\)
give the inverse inertia tensors $\lr(\s)$ of the twisted sectors in terms of the inverse inertia tensor $L$ of each $H$-invariant CFT.

The paper is organized as follows.  Our motivation for this development is given in Sec.~\ref{Jansec}, where we find the \textit{first duality transformation} which identifies all the solutions of the OVME as sectors of the orbifolds
\(
\ad \sp \zl \subset \dl \subset Aut(g) \sp \gh=\oplus_{J=0}^{\l-1}\gb^J \sp \gb^J \cong \gb. \label{introg}
\)
Here the cyclic group $\zl$ and the dihedral group $\d_\l$ permute the copies $\{\gb^J\}$, and $A(\dl)$ is any $\dl$(permutation)-invariant CFT on $g$.

The technical center of the paper is found in Secs.~\ref{mainsec} and \ref{covmesec}.  In Sec.~\ref{mainsec}, we generalize the result of Sec.~\ref{Jansec} to find a set of duality transformations which constructs the stress tensors of all the sectors of the orbifolds
\(
\az\supset\ad.
\)
In these cases, the untwisted sector of each orbifold is described by the stress tensor  
\(
T(z)=\sum_{J,L=0}^{\l-1}L_{J-L}^{ab} :J_{aJ}(z)J_{bL}(z): 
\sp  a,b=1,...,\dg
\) 
of any $\zl$(permutation)-invariant CFT on $g$.  Here, $J_{aJ}$ are $\l$ copies of currents at level $k$ and $L_{J-L}^{ab}$ is any solution of the reduced VME of the $\z_\l$(permutation)-invariant CFT's (see Subsec.~\ref{CVMEsec}).  The stress tensors $\hat{T}_\s$ of the twisted sectors $\s$ of the orbifolds \taz\ are then obtained by the duality transformations in terms of the solution $L_{J-L}^{ab}$ as
\namegroup{introTgroup}
\alpheqn
\(
\hat{T}_\s (z)=\sum_{r=0}^{\r(\s)-1}\sum_{j,l=0}^{\lors-1}\lr_r^{a(j)b(l)}(\s) :\hat{J}_{aj}^{(r)}(z) \hat{J}_{bl}^{(-r)}(z): \sp \s=1,...,\l-1
\)
\(
\lr_r^{a(j)b(l)}(\s)=\frac{1}{\r(\s)}\sum_{s=0}^{\r(\s)-1}e^{-\frac{\tp N(\s)rs}{\r(\s)}}L^{ab}_{\lors s+j-l}\ \label{intromap}
\)
\(
\gb_{\r(\s)}=\oplus_{j=0}^{\lors-1}\gb_{\r(\s)}^j \label{introgrho}
\)
\reseteqn
where $\hat{J}_{aj}^{(r)}$ satisfies the orbifold affine algebra $\gb_{\r(\s)}$.  Here $\r(\s)$ is the order of the orbifold affine algebra and also the order of the element $h_\s$ of $\z_\l$.  The integers $N(\s)$ are defined in the text.

The explicit form of the new duality transformations in (\ref{intromap}) is a central result of this paper, and, in this case, the duality transformations are seen to be discrete Fourier transforms.  Using (\ref{introTgroup}) and the representation theory of orbifold affine algebra$^{\rf{9}}$, we compute in Eq.~(\ref{OVMEcw}) the ground state conformal weights $\hat{\Delta}_0(\s)$ of all the sectors of any orbifold \taz.  In Sec.~\ref{covmesec} we also find the \textit{cyclic orbifold Virasoro master equation}, which contains all the sectors of the orbifolds \taz.

Extending the development of Sec.~\ref{mainsec}, general orbifolds of the form
\(
\frac{A(H)}{H} \sp H\subset Aut(g)
\)
are discussed for each fixed $H$ in Sec.~\ref{gensec}, where $A(H)$ is any $H$-invariant CFT on the ambient Lie algebra $g$.  Here the symmetry group $H$ can be any finite subgroup of inner or outer automorphisms of $g$, which itself may be simple or semisimple.  Formally, $H$ can be extended to Lie groups.  Our results in this section include the form of the \textit{general twisted current algebra} for all $H \subset Aut(g)$ and the \textit{general duality transformations} (which include (\ref{introTgroup})) for the stress tensors of the twisted sectors of any orbifold $A(H)/H$.  We also give the $A(H)/H$ \textit{orbifold Virasoro master equation} which collects the inverse inertia tensors of each sector $\s$ of all the orbifolds $A(H)/H$ at fixed $H\subset Aut(g)$.  The general formulation is complete up to the solution of a well-defined eigenvalue problem for each symmetry group.  The representation theory of the general twisted current algebra is apparently a synthesis of the representation theories of previously-known cases.

The explicit form of the stress tensors of all the permutation orbifolds $A(S_3)/S_3$ is obtained in Subsec.~\ref{S3sec} and simple examples in \taz\ are discussed in Subsec.~\ref{Gdiagsec} and App.~E.  A number of conceptual matters are also noted in App.~E, including ``copy'' orbifolds versus ``interacting'' orbifolds and the ``uncertainty relations'' associated to the duality transformations.

\newsection{Solution of the OVME Identification Puzzle}
\label{Jansec}
In this section, the OVME identification puzzle is reviewed and solved.  The central observation here is the first duality transformation, which maps the solutions of the OVME onto a particular set of solutions of the VME.
\subsection{Orbifold affine algebra}
We begin with the \textit{orbifold affine algebra}$^{\rf{9}}$ $\gb_\l$
\alpheqn 
\(
\hat{J}^{(r)}_a(z)\hat{J}^{(s)}_b(w)=
\frac{\hat{G}_{ab}\delta_{r+s,0\hspace{.05in}mod\hspace{.04in}\lambda}}{(z-w)^2}
+\frac{if_{ab}^{ \ \ c}\hat{J}_c^{(r+s)}(w)}{(z-w)} + O((z-w)^0) \label{OJJOPE}
\)
\(\hat{J}^{(r)}_a(z)=\sum_{m\in\sz}\hat{J}^{(r)}_a(m+\frac{r}{\lambda}) \ z^{-1-m-\frac{r}{\lambda}} \sp \hat{J}_a^{(r)}(ze^{\tp})= e^{-\frac{\tp r}{\l}}\hat{J}_a^{(r)}(z)
\) 
\(
\hat{J}^{(r\pm \l)}(z)=\hat{J}^{(r)}(z) \hspace{.3in} (\textup{periodicity convention}) \label{Jtwistperiod}
\)
\(
\hat{J}_a^{(r)}(m+\frac{r}{\lambda})|0\rangle=0\textrm{\ \ when\ \ }(m+\frac{r}{\lambda})\geq 0 \label{jvac}
\)
\(
\arange\sp r,s=0,...,\l-1\sp m\in\z
\)
\reseteqn
\noindent
where the \textit{order} $\lambda$ of $\gb_\l$ is any positive integer and $\gb_{\l=1}$ is affine Lie algebra$^{\rf{10}-\rf{12}}$.  The state $|0\rangle$ is the ground state of the algebra.  The quantities $f_{ab}^{\ \ c}$ are the structure constants of any \textit{semisimple} Lie algebra $\gb$ and the metric $\hat{G}_{ab}$ is
\(
\hat{G}_{ab}=\stackrel{S-1}{\pa} \hspace{-.1in} \hat{k}_\a \eta_{ab}^{\a} \sp \hat{k}_\a=\l k_\a \sp \gb=\oplus_{\a=0}^{S-1} \gb^\a\label{semia}
\)
when $\gb$ has $S$ generically distinct simple components $\gb^\a$, $\a=0,...,S-1$.  Here $\eta^\a_{ab}$ is the Killing metric of $\gb^\a$ and unitarity on compact $\gb$ requires that the invariant levels $\hat{x}_\a$ satisfy
\(
\hat{x}_\a=\l x_\a\sp x_\a=\frac{2 k_\a}{\psi_\a^2} \in \z^+ \label{xl}
\) 
where $\psi_\a$ is the highest root of $\gb^\a$.  The metric in (\ref{semia}) assigns the same order $\l$
\(
\gb_\l \equiv \oplus_{\a=0}^{S-1} \gb_{\l}^\a \label{semib}
\)
to each component $\gb_{\l}^\a$ of the orbifold affine algebra.

The representation theory of orbifold affine algebras (including the ground state property (\ref{jvac}) and the unitarity statement (\ref{xl})) was obtained via the \textit{orbifold induction procedure} in Ref.~\rf{9}, to which the reader is referred for further detail.  It was also argued in Ref.~\rf{9} that orbifold affine algebras occur in the twisted sectors of cyclic orbifolds, and a more conventional derivation of orbifold affine algebra (indicated but not completed in Ref.~\rf{9}) is given in Sec.~\ref{mainsec}.
\subsection{The OVME}
Recently, a large class of conformal stress tensors$^{\rf{5}}$
\namegroup{OVMEgrp}
\alpheqn
\(
\hat{T}(z)\hat{T}(w)=\frac{\hat{c}/2}{(z-w)^4}+\frac{2\hat{T}(w)}{(z-w)^2}+\frac{\partial_w\hat{T}(w)}{(z-w)}+O((z-w)^0)
\) 
\(
\hat{T}(z)=\sum_{r=0}^{\lambda-1}\lr_r^{ab}:\hat{J}_a^{(r)}(z)\hat{J}_b^{(-r)}(z): \label{stresstensor}
\)
\(
\lr_r^{ab} =  2 \lr_r^{ac}  \hat{G}_{cd} \lr_r^{db} - \sum^{\lambda-1}_{s=0} \lr_s^{cd} [\lr_{r+s}^{ef} f_{ce}^{\ \ a}f_{df}^{\ \ b} + f_{ce}^{\ \ f}f_{df}^{\ \ (a} \lr_r^{b) e}]\label{OVME}
\)
\(
\lr_r^{ab}=\lr_r^{ba} \comment{($a\lra b$ symmetry)} \label{ovmesym}
\)
\(
\lr_{r \pm \lambda}^{ab}=\lr_{r}^{ab}\comment{(periodicity)}\label{period}
\)
\getletter{simpleovmelett}
\(
\lr_r^{ab}=\lr_{\lambda-r}^{ab} \comment{(gauge condition)}\label{gcond}
\)
\(
\hat{c}=2 \hat{G}_{ab} \sum_{r=0}^{\lambda-1}\lr_r^{ab} \sp
\hat{\Delta}_0 = \hat{G}_{ab} \sum_{r=0}^{\lambda -1} \lr_r^{ab} \frac{r (\lambda - r)}{2 \lambda^2}\label{delta} 
\) 
\reseteqn
was constructed on the orbifold affine algebra $\gb_\l$, where (\ref{OVME}-\ref{simpleovmelett}) is called the \textit{orbifold Virasoro master equation} (OVME).  The operator-product normal ordering 
\(
:A(w)B(w):=\oint_w\frac{dz}{\tp}\frac{A(z)B(w)}{z-w}
\)
is used throughout this paper, where the contour does not encircle the origin.

In (\ref{OVMEgrp}), the set of coefficients $\lr$, called collectively the inverse inertia tensor of $\hat{T}$, is any solution of the OVME.  The quantity $\hat{c}$ is the central charge of any solution $\lr$ and $\hat{\Delta}_0$ is the conformal weight of the ground state
\(
\hat{T}(z)=\sum_{m\in\sz}L(m)z^{-m-2}\sp L(m\geq 0)|0\rangle=\delta_{m, 0}\hat{\Delta}_0|0\rangle .
\)
The \textit{symmetries} (\ref{ovmesym}), (\ref{period}) and (\ref{gcond}) of the solutions of the OVME will play an important role in the present paper and we will sometimes need combinations of these such as
\(
\lr_r^{ab}=\lr_{-r}^{ab}\sp r=0,...,\l-1 \label{combo}
\)
which combines the gauge condition with periodicity.

The OVME is a large algebraic system, consisting of
\(
n_D(\gb,\l)=(\hl+1)\frac{\dg(\dg+1)}{2} \label{nD}
\)
coupled quadratic equations for the same number of independent unknowns $\lr$.
$\lfloor x\rfloor$ is the integer less than or equal to $x$.  The number of physically inequivalent solutions to the OVME is therefore expected to be
\(
N_D(\gb,\l)=2^{n_D({\sgb},\l) - \mbox{\scriptsize{\textup{dim}}}\sgb} \label{ND}
\)
for each set of levels \{$x_\a$\} of $\gb_\l$.  Many properties and new solutions of the OVME were discussed in Ref.~\rf{5}, to which we refer the reader for further detail.

Two simple facts about the OVME will be important for our development here.  First, the OVME contains the Virasoro master equation$^{\rf{1}-\rf{3}}$ (VME) when $\l=1$,
\(
\l=1:\ \ \textup{OVME\ $\rightarrow$\ VME}
\)
and in this case the conformal weight $\hat{\Delta}_0 \rightarrow \Delta_0$ of the ground state is zero.  Other relevant properties of the VME will be reviewed in Subsec.~\ref{VMEsec}.  Second, the result (\ref{delta}) shows that the ground state conformal weight of the generic solution of the OVME is nonvanishing,
\(
\l\geq 2:\ \ \hat{\Delta}_0\neq 0
\)
which indicates that each solution of the OVME for $\l\geq 2$ gives a stress tensor of a twisted sector of an orbifold.  Except in special cases, however, these orbifolds were not identified in Ref.~\rf{5}.  This is the OVME identification puzzle.

In what follows we find a new duality transformation which identifies the orbifolds of the OVME. 

\subsection{The First duality transformation} \label{fdt}
For this analysis, it is convenient to introduce a more explicit notation,
\namegroup{OVMEgf}
\alpheqn
\(
a,b \rightarrow a(\a),b(\b)\sp \hat{G}_{ab}\rightarrow\hat{G}_{a(\a)b(\b)}
\)
\(
\a,\b=0,...,S-1 \sp a(\a)=1,...,\textup{dim} \gb^\a, \ b(\b)=1,...,\textup{dim} \gb^\b
\)
\(
\hat{G}_{a(\a)b(\b)}=\hat{G}_{ab}^{(\a)} \delta_{\a \b}\sp
\hat{G}_{ab}^{(\a)}= \l G_{ab}^{(\a)}  \sp G_{ab}^{(\a)}=k_\a\eta_{ab}^\a
\)
\(
f_{a(\a)b(\b)}^{\ \ \ \ \ \ \ c(\gamma)}=f_{ab}^{(\a)\ c}\delta_{\a \b}\delta_{\b}^{\ \gamma}
\)
\reseteqn
for the semisimplicity specified by the composite indices $a,b$ in (\ref{semia}) and (\ref{semib}).  The symmetries (\ref{ovmesym}), (\ref{period}) and (\ref{gcond}) of the solutions of the OVME take the form
\alpheqn
\begin{eqnarray}
\lr_r^{a(\a) b(\b)}&=&\lr_r^{b(\b) a(\a)} \hspace{.3in} (a(\cdot) \leftrightarrow b(\cdot)) \\
&=&\lr_{r \pm \l}^{a(\a) b(\b)} \hspace{.3in} (\textup{periodicity}) \\
&=&\lr_{\l-r}^{a(\a) b(\b)} \hspace{.3in} (\textup{gauge condition}) 
\end{eqnarray}
\reseteqn
in the explicit notation.

The \textit{first duality transformation} is the discrete Fourier transform $L$ of any solution $\lr$ of the OVME,
\vspace{-.3in}
\namegroup{FTgroup}
\alpheqn
\begin{eqnarray}
L_{J-L}^{a(\a) b(\b)} & \equiv & \sum_{r=0}^{\l-1} e^{\frac{2 \pi i (J-L) r}{\l}} \lr_r^{a(\a) b(\b)} \label{theSSmap} \\
L_{J-L\pm\l}^{a(\a) b(\b)} &=& L_{J-L}^{a(\a) b(\b)}\comment{(periodicity)}
\label{lper}
\end{eqnarray}
\(
J,L=0,...,\l-1
\)
\reseteqn
and its inverse (see Eq. (\ref{period}))
\(
\lr_r^{a(\a) b(\b)}=\frac{1}{\l} \sum_{K=0}^{\l-1} e^{-\frac{2 \pi i K r}{\l}} L_K^{a(\a) b(\b)} \label{Janinv}
\)
which follows from the identity (\ref{perid}).  The Fourier transform inherits the symmetries
\namegroup{inherits}
\alpheqn
\begin{eqnarray}
L_{J-L}^{a(\a) b(\b)} &=& L_{J-L}^{b(\b) a(\a)}\comment{(from $a(\cdot) \leftrightarrow b(\cdot)$ of OVME )} \\
&=& L_{L-J}^{a(\a) b(\b)}\comment{(from gauge condition and periodicity of OVME)}
\end{eqnarray}
\reseteqn
which, together with the periodicity condition (\ref{lper}), will play an important role in the discussion below.

We now consider the Fourier transform of each term in the OVME, for example
\alpheqn
\(
\sum_{r,\gamma}e^{\frac{\tp (J-L)r}{\l}}\lr_r^{a(\a)c(\gamma)}\hat{G}_{cd}^{(\gamma)}\lr_r^{d(\gamma)b(\b)}
=\frac{1}{\l^2}\sum_{r,M,N,\gamma}e^{\frac{\tp (J-L+M+N)r}{\l}}L_M^{a(\a)c(\gamma)}\hat{G}_{cd}^{(\gamma)}L_N^{d(\gamma)b(\b)}
\)
\(
=\frac{1}{\l}\sum_{M,N,\gamma}\delta_{J-L+M+N,0\mod\l}L_M^{a(\a)c(\gamma)}\hat{G}_{cd}^{(\gamma)}L_N^{d(\gamma)b(\b)}
=\sum_{M,\gamma}L_{L-J-M}^{a(\a)c(\gamma)}G_{cd}^{(\gamma)}L_M^{d(\gamma)b(\b)}
\) \label{peridone}
\reseteqn
where we have used Eqs. (\ref{period}) and (\ref{perid}).  Collecting all the terms, we find the dual form of the OVME:
\vspace{-.2in}
\namegroup{dualsys}
\alpheqn
\begin{eqnarray}
L_{J-L}^{a(\a) b(\b)} & = & 2\sum_{M=0}^{\l-1}\sum_{\gamma=0}^{S-1} L_{J-L-M}^{a(\a) c(\gamma)} G_{cd}^{(\gamma)} L_M^{d(\gamma) b(\b)} - L_{J-L}^{c(\a) d(\b)} L_{J-L}^{e(\a) f(\b)} f_{ce}^{(\a)\ a} f_{df}^{(\b)\ b} \nonumber \\
&\ & - L_0^{c(\a)d(\a)} f_{ce}^{(\a)\ f} f_{df}^{(\a) \ a} L_{J-L}^{b(\b)e(\a)}
- L_0^{c(\b)d(\b)} f_{ce}^{(\b)\ f} f_{df}^{(\b) \ b} L_{J-L}^{a(\a)e(\b)}
 \label{CVME}
\end{eqnarray}
\(
L_{J-L}^{a(\a)b(\b)}=L_{J-L}^{b(\b)a(\a)}=L_{L-J}^{a(\a)b(\b)} \label{CVMEsym}
\)
\(
a(\a)=1,...,\dg^\a\sp \a,\b=0,...,S-1\sp J,L=0,...,\l-1. \label{notsym}
\)
\reseteqn
Similarly, we find for the dual form of the central charge
\(
\hat{c}= 2\sum_{\a,\b=0}^{S-1} \hat{G}_{a(\a)b(\b)} \sum_{r=0}^{\l-1} \lr_r^{a(\a)b(\b)} = 2 \sum_{\a=0}^{S-1}\hat{G}_{ab}^{(\a)} \sum_{r=0}^{\l-1} \lr_r^{a(\a) b(\a)}= 2 \l \sum_{\a=0}^{S-1} G_{ab}^{(\a)} L_0^{a(\a)b(\a)} \label{cone}
\) 
where we have used (\ref{Janinv}) and (\ref{perid}) to obtain the last form of $\hat{c}$.  

Including the symmetries (\ref{CVMEsym}) of $L$, the dual form of the OVME is seen to have the same number of equations and unknowns $n_D$ (and hence the same number of solutions $N_D$) as those given for the OVME in (\ref{nD}) and (\ref{ND}).

In the following subsection, we will identify the dual system (\ref{dualsys}), (\ref{cone}) as a certain consistent subansatz of the Virasoro master equation.

\subsection{The $\subsecd_\l$(permutation)-invariant CFT's in the VME} \label{VMEsec}
We consider next the general affine-Virasoro construction$^{\rf{1}-\rf{3}}$
\namegroup{VME}
\alpheqn 
\(
T(z)T(w)= \frac{c/2}{(z-w)^4}+ \frac{2 T(w)}{(z-w)^2}+\frac{\partial_w T(w)}{(z-w)} + O((z-w)^0)
\)
\(
T(z)=L^{ab}:J_a(z)J_b(z):\sp a,b=1,..,\dngg \label{VMEaa}
\)
\(
L^{ab} =  2 L^{ac} G_{cd} L^{db} - L^{cd} L^{ef} f_{ce}^{\ \ a}f_{df}^{\ \ b} - L^{cd} f_{ce}^{\ \ f}f_{df}^{\ \ (a}L^{b) e} \label{VMEa}
\)
\(
L^{ab}=L^{ba}\comment{($a\lra b$ symmetry)}
\)
\(
c=2G_{ab}L^{ab} \label{gabdef}
\)
\reseteqn
on the currents $J$ of the affine Lie algebra$^{\rf{10}-\rf{12}}$ on $g$
\alpheqn 
\(
J_a(z)J_b(w)=
\frac{G_{ab}}{(z-w)^2} + \frac{if_{ab}^{ \ \ c}J_c(w)}{(z-w)} + O((z-w)^0) \label{JJOPE}
\)
\(
J_a(ze^{\tp})= J_a(z).
\)
\reseteqn
Here (\ref{VMEa}) is the \textit{Virasoro master equation} (VME) and $L$, called the inverse inertia tensor of $T$, is any solution of the VME.

With an eye to the previous subsection, we choose here the specific \textit{semisimplicity}:
\alpheqn
\(
\gh \equiv \oplus_{I=0}^{\l-1} \gb^I\sp \gb^I= \oplus_{\a=0}^{S-1}\gb^{\a I}\cong \gb
\label{gprime}
\)
\(
a,b\rightarrow a(\a J),b(\b L)\sp
G_{ab}\rightarrow G_{a(\a J)b(\b L)}
\)
\(
G_{a(\a J)b(\b L)}=G_{a(\a)b(\b)}\delta_{JL}\sp 
G_{a(\a)b(\b)}=G_{ab}^{(\a)}\delta_{\a \b}\sp
G_{ab}^{(\a)}=k_\a \eta_{ab}^\a \label{VMEssg}
\)
\(
f_{a(\a J)b(\b L)}^{\ \ \ \ \ \ \ \ \ \ c(\gamma M)}=f_{ab}^{(\a)\ c}\delta_{\a \b}\delta_{\a}^{\ \gamma}\delta_{JL}\delta_{J}^{\ M} \label{VMEssf}
\)
\(
\a,\b,\gamma=0,...,S-1\sp I,J,L,M=0,...,\l-1
\)
\reseteqn
where Greek letters $\a,\b,...$ label $S$ generically distinct simple components and capital letters $I,J,...$ label $\l$ copies of the same Lie algebra $\gb^I\cong \gb$ defined in (\ref{semia}).  The quantities $G_{ab}^{(\a)}$ and $f_{ab}^{(\a)\ c}$ in (\ref{VMEssg}) and (\ref{VMEssf}) are the metric and structure constants of the distinct components, and we emphasize that these quantities are exactly those defined in (\ref{OVMEgf}).

We focus next on the {\textit{$\dl$(permutation)-invariant CFT's}}\footnote{The $H$-invariant CFT's on $\gh$ are discussed in Refs.~\rf{4}, \rf{47} and \rf{3}:  For any $H \subset Aut(g)$, the $H$-invariant CFT's $A(H)$ on $g$ are described by a consistent subansatz of the VME, i.e. a reduced VME with an equal number of equations and unknowns.  It is also known that $A(H_1) \supset A(H_2)$ when $H_1 \subset H_2 \subset Aut(g)$.} on $g$, which are all the solutions of the VME with a dihedral symmetry $\dl$ that permutes the copies of the algebra $\gb$:  
\alpheqn
\begin{eqnarray}
L^{a(\a,J)b(\b,L)} &=& L^{b(\b,L)a(\a,J)}  \comment{$(a(\cdot) \leftrightarrow
 b(\cdot))$} \label{ssLab}\\
&=& L^{a(\a,J \pm \l)b(\b,L)} = L^{a(\a,J)b(\b,L \pm \l)}  \comment{(periodicity})\\
&=& L^{a(\a,L)b(\b,J)}\hsp{.84} \comment{(reflection: $s\in\d_\l$)} \label{disymm} \\
&=& L^{a(\a,J+1)b(\b,L+1)}\hsp{.25} \comment{($\z_\l$ (in $\d_\l$) invariance)}. \label{VMESScyclic} 
\end{eqnarray}
\reseteqn
The solution of the cyclic condition (\ref{VMESScyclic}) is 
\alpheqn
\begin{eqnarray}
L^{a(\a,J)b(\b,L)} &=& L_{J-L}^{a(\a) b(\b)} \\
 &=& L_{L-J}^{b(\b) a(\a)} \comment{$(a(\cdot) \leftrightarrow b(\cdot))$} \\
 &=& L_{J-L \pm \l}^{a(\a) b(\b)} \comment{(periodicity)} \\
 &=& L_{L-J}^{a(\a) b(\b)} \comment{(reflection)}
\end{eqnarray}
\reseteqn
and we recognize these symmetries as identical to those of the Fourier transform $L$ in (\ref{FTgroup}) and (\ref{inherits}).

Moreover, using the VME (\ref{VMEa}) to work out the reduced VME of the $\dl$(permutation)-invariant CFT's, we find exactly the dual form of the OVME in (\ref{dualsys})!

This allows us to identify the Fourier transform $L$ of any particular solution $\lr$ of the OVME as the inverse inertia tensor $L$ of a particular $\dl$(permutation)-invariant CFT.

Finally, we may rearrange the central charge (\ref{gabdef}) to find that
\(
c=2 G_{a(\a J)b(\b L)} L^{a(\a J)b(\b L)} = 2\l \sum_\a G_{ab}^{(\a)} L_0^{a(\a)b(\a)}=\hat{c}
\)
where we have used (\ref{VMEssg}) and (\ref{cone}).  This relation tells us that the central charge $c$ of the $\dl$(permutation)-invariant CFT is the same as the central charge $\hat{c}$ of the solution of the OVME.

Our interpretation\footnote{As we will discuss in a sequel, the sectors described by the OVME also occur in the permutation orbifolds $A(\d_\l)/\d_\l$.} of these phenomena is that \textbf{the first duality transformation in (\ref{FTgroup}) and (\ref{Janinv})}:
\(
\lr \hspace{.1in} \leftrightarrow \hspace{.1in} L
\)
\textbf{identifies each solution $\mathbf{\lr}$ of the OVME as the inverse inertia tensor of a twisted sector of the orbifold}
\(
\ad \sp \zl \subset \dl \subset Aut(g) \sp \gh=\oplus_{J=0}^{\l-1}\gb^J \sp \gb^J \cong \gb 
\)
\textbf{where A($\dl$) is the $\dl$(permutation)-invariant CFT specified by the solution $L$ of the VME}.  The following section provides an alternate and more general derivation of this interpretation in the context of standard orbifold theory.  The alternate derivation will also construct the stress tensors of all the other twisted sectors of each orbifold \tad, beyond the single sector $\hat{T}$ in (\ref{stresstensor}).

For reference below we also introduce the following partially composite notation:
\alpheqn
\(
a(\a J),b(\b L)\rightarrow a(J),b(L)
\)
\(
G_{a(\a J)b(\b L)}\rightarrow G_{a(J)b(L)}= G_{ab}\p \delta_{JL} \sp
G_{a(\a)b(\b)}\rightarrow G_{ab}\p \equiv \stackrel{\ }{\pan} \hspace{-.05in} k_\a\eta_{ab}^\a
\)
\(
f_{a(\a J)b(\b L)}^{\ \ \ \ \ \ \ \ \ \ c(\gamma M)}\rightarrow f_{ab}^{\ \ c}\delta_{JL}\delta_L^{\ M}
\)
\(
L^{a(\a J)b(\b L)}\rightarrow L^{a(J)b(L)}=L^{ab}_{J-L}\sp
J_{a(\a J)}\rightarrow J_{a(J)}=J_{aJ}
\)
\reseteqn
which suppresses the semisimplicity labeled by $\a,\b$ or $\gamma$ but leaves explicit the semisimplicity of the copies labeled by $J,L$ or $M$.

In this notation, the $\dl$(permutation)-invariant CFT's take the form
\namegroup{DVME}
\alpheqn
\(
T(z)=\sum_{J,L=0}^{\l-1} L_{J-L}^{ab}:J_{aJ}(z)J_{bL}(z):
\)
\getletter{DVMEeq}
\(
L_{J-L}^{a b} =  2\sum_{M=0}^{\l-1} L_{J-L-M}^{ac} G_{cd}\p L_M^{db} - L_{J-L}^{c d} L_{J-L}^{e f} f_{ce}^{\ \ a} f_{df}^{\ \ b} - L_0^{cd} f_{ce}^{\ \ f} f_{df}^{\ \ (a} L_{J-L}^{b) \ e} \label{CVMEcomp}
\)
\getletter{DVMEsymm}
\(
L_{J-L}^{ab}=L_{J-L}^{ba}=L_{L-J}^{ab}=L_{J-L \pm \l}^{ab}
\)
\(
c=2\sum_{J,L=0}^{\l-1} G_{a(J)b(L)}L^{a(J)b(L)}=2\l  G_{ab}\p L_0^{ab}.
\)
\(
a,b=1,...,\dg\sp \gb^I \cong \gb \sp J,L=0,...,\l-1.
\)
\reseteqn
This system retains the numbers $n_D$ and $N_D$ of unknowns, equations and solutions given in (\ref{nD}) and (\ref{ND}).  We will also need the first duality transformation
\(
L_{J-L}^{ab} \equiv \sum_{r=0}^{\l-1}  e^{\frac{2 \pi i (J-L) r}{\l}} \lr_r^{a b}, \hspace{.3in} \lr_r^{ab}=\frac{1}{\l} \sum_{K=0}^{\l-1} e^{-\frac{2 \pi i K r}{\l}} L_K^{ab} \label{simpcase}
\)
in this notation.

\newsection{\hspace{-.1in}The orbifolds of the $\mz_\l$(permutation)-invariant CFT's}
\label{mainsec}
In this section we develop a synthesis of standard\footnote{At the level of examples, the algebraic approach$^{\rf{28}-\rf{32}}$ to orbifold theory predates the geometric approach$^{\rf{33}-\rf{36}}$.  In particular, twisted scalar fields, a twisted ``copy'' of a chiral algebra in a twisted sector, and coupling to the untwisted sector were noted as early as 1971 in Sec.~4 of Ref.~\rf{28}.} orbifold theory (see Refs.~\rf{28}-\rf{42}, and in particular Subsec.~3.8 of Ref.~\rf{9}) and the theory of the $H$-invariant CFT's on $\gh$ (see Refs.~\rf{4}, \rf{47} and \rf{3})  for the case $H=\z_\l$(permutation).  In particular, we use this development to find a set of duality transformations which gives the stress tensors of all the sectors of every orbifold 
\(
\az
\)
where $A(\z_\l)$ is any $\z_\l$(permutation)-invariant CFT.  The results include as special cases all the sectors of all the orbifolds 
\(
\ad\subset\az\sp \dl(\textup{permutation}) \supset\z_\l(\textup{permutation})
\)
associated to the OVME.  General orbifold theory $A(H)/H$, $H\subset Aut(g)$ is considered along the same lines in Sec.~\ref{gensec}.

\subsection{$\subsecz_\l$ automorphisms}
We consider first the $\z_\l$ automorphisms of $\lambda$ copies $\{\gb^I\}$ of a simple algebra $\gb$ and their corresponding currents $J_{aI}\equiv J_{a(I)}$, where each copy is taken at affine level $k$,
\namegroup{cycdefs}
\alpheqn
\(
\gh=\oplus_{I=0}^{\l-1}\gb^I \sp \gb^I \cong \gb
\)
\(
G_{a(J)b(L)}=G_{ab}\delta_{JL}\sp G_{ab} = k \eta_{ab} \sp f_{a(J)b(L)}^{\ \ \ \ \ \ \ \ c(M)}=f_{ab}^{\ \ c}\delta_{JL}\delta_{L}^{\ M} \label{cycgf}
\)
\(
J_{aJ}(z)J_{bL}(w)=\delta_{JL}[\frac{k\eta_{ab}}{(z-w)^2}+\frac{if_{ab}^{\ \ c}J_{cJ}(w)}{(z-w)}]+\reg
\)
\(
\arange\sp
I,J,L,M=0,...,\l-1.
\)
\reseteqn
This corresponds to the case $S=1$ in Subsec.~\ref{VMEsec}, and the generalization to arbitrary $S$ is discussed in Sec.~\ref{covmesec}.  The elements of $\z_\l$ are
\namegroup{hinaut}
\alpheqn
\(
h_\sigma  \in\z_\l\subset Aut(g), \hspace{.3in} h_\sigma  h_{\sigma'}=h_{\sigma+\sigma'\ mod \ \l}
\)
\(
\sigma=0,...,\l-1 
\)
\reseteqn
where $h_0$ is the identity element.  Let $\ws$ be the regular representation of $h_\sigma$ so that
\alpheqn
\(
J_{aJ}(z)\p=\ws_{JL}J_{aL}(z)\sp\ws\in\z_\l \subset \textup{Aut}(g)
\)
\(
\ws_{JL}=\delta_{J+\s,L\mod\l}\hsp{.6}  \label{wdef}
\)
\reseteqn
is an (outer) automorphism of the affine algebra.

For the discussion below, we will need some definitions.  Let $\r(\sigma)$ be the \textit{order} of $h_\sigma$ and $\ws$, so that $\r(\sigma)$ is the smallest positive integer satisfying
\(
\ws^{\r(\s)}=1. \label{defrho}
\)
It is known that $\r(\l-\s)=\r(\s)$, and we also define two integers $M(\s)$ and $N(\s)$ for each element $h_\sigma$
\vspace{-.2in}
\namegroup{defints}
\alpheqn
\(
M(\s)=\frac{\s\r(\s)}{\l} \label{added}
\)
\(
\ \ \ \ \ \ \ \ \ \ \ \ \ N(\sigma)M(\sigma)=1\rmod\r(\s)
\)
\reseteqn
where we choose the conventions
\(
N(0)=0; \hspace{.3in} 1\leq N(\sigma) \leq \r(\s)-1 \sp \s=1,...,\l-1.  \label{Nconv}
\)
As an example, note that $\r(1)=\l$ and hence $M(1)=N(1)=1$.  Other examples and a closed form for $N(\s)$ are given in App.~B.

\subsection{The $\subsecz_\l$(permutation)-invariant CFT's} \label{CVMEsec}
The set $A(\z_\lambda)$ of $\z_\lambda$(permutation)-invariant conformal field theories is defined to include all inverse inertia tensors $L$ in the VME which are invariant under all the $\z_\l$ automorphisms in (\ref{hinaut}):
\alpheqn
\(
L^{a(J\p)b(L\p)} \ \ws_{J\p J}\ \ws_{L\p L}=L^{a(J)b(L)} \sp \sigma=0,..,\l-1
\label{LwwL}
\)
\(
L^{a(J)b(L)}=L^{b(L)a(J)}\comment{($a(\cdot)\lra b(\cdot)$ symmetry)}.
\label{ssLL}
\)
\reseteqn
The solution of (\ref{LwwL}) and (\ref{ssLL}) is
\namegroup{Lcyc}
\alpheqn
\(
L^{a(J)b(L)}=L^{a(J+1)b(L+1)}=L_{J-L}^{ab}\comment{($\z_\l$ invariance)}
\)
\(
L_{J-L}^{ab}=L_{L-J}^{ba}\comment{$(a(\cdot)\lra b(\cdot)$ symmetry)}
\label{zlsyms}
\)
\reseteqn
and we will also specify the periodicity
\(
L_{J-L\pm \l}^{ab}=L_{J-L}^{ab}. \label{zlper}
\)
Note that these specifications differ from the $\dl$(permutation)-invariant CFT's $A(\dl)$ because (\ref{Lcyc}) and (\ref{zlper}) do not include the reflection symmetry (\ref{disymm}).  Of course
\(
A(\dl)\subset A(\z_\l) 
\)
because the solutions with the extra condition (\ref{disymm}) are included as special cases of the $\z_\l$(permutation)-invariant CFT's.

The reduced system which describes the set of all $\z_\l$(permutation)-invariant CFT's
\namegroup{redVME}
\alpheqn
\(
T(z)=\sum_{J,L=0}^{\l-1} L_{J-L}^{ab}:J_{aJ}(z)J_{bL}(z): \label{redT}
\)
\getletter{redlet}
\ba
L_{J-L}^{ab}&=&2k\sum_{M=0}^{\l-1} L_{J-L-M}^{ac}\eta_{cd}L_M^{db}-L_{J-L}^{cd}L_{J-L}^{ef}f_{ce}^{\ \ a}f_{df}^{\ \ b}\nonumber\\
& &-L_0^{cd}f_{ce}^{\ \ f}[f_{df}^{\ \ a}L_{L-J}^{be}+f_{df}^{\ \ b}L_{J-L}^{ae}] \label{redcVME}
\ea
\getletter{symlet}
\(
L^{ab}_{J-L}=L_{L-J}^{ba}=L^{ab}_{J-L\pm\l}  \label{redVMEsym}
\)
\(
c=2\l k \eta_{ab}L^{ab}_0 \label{cLzero}
\)
\(
\arange\sp J,L=0,...,\l-1 
\)
\reseteqn
then follows from the general affine-Virasoro construction (\ref{VME}).  

Using the symmetries in (\ref{redVMEsym}), one finds that the inverse inertia tensors can be pulled back into the fundamental range
\alpheqn
\(
L_K^{ab}, \ 1\leq K\leq\lfloor \frac{\l-1}{2} \rfloor, \ \ \forall\  a,b;\hsp{.3} L^{ab}_0,\ \ \forall\  a \leq b
\)
\(
L_{\frac{\l}{2}}^{ab}, \ \forall\  a\leq b \ \textup{when $\l$ is even}.
\)
\reseteqn
It follows that the reduced VME (\ref{redVME}\ \ref{redlet},\ref{symlet}) is a consistent system with 
\(
n_C(\gb,\l)=\dg \{\frac{\l(\dg-1)}{2}+\hl+1 \} \sp \gb^I\cong \gb \label{nc}
\)    
equations and unknowns, and hence the generically-expected number of $\z_\l$(permutation)-invariant CFT's is 
\(
N_C(\gb,\l)=2^{n_C({\sgb},\l)-\mbox{\scriptsize{\textup{dim}}}\sgb} \label{NCeq}
\)
on each level $k$ of $\gb$.  The numbers $N_C$ in (\ref{NCeq}) and $N_D$ in (\ref{ND}) agree for $\l=2$ because $\z_2 \cong \d_2$.  Numerically one finds for example that 
\alpheqn
\(
N_C(SU(2),\lambda=2)= 512\sp N_C(SU(2),\lambda=3)=4096
\)
\(
N_C(SU(3),\lambda=2)\approx 18 \ \textrm{quintillion}
\)
\reseteqn
and the number of $\z_\l$(permutation)-invariant CFT's increases exponentially with increasing $\l$ and/or $\dg$.

The reduced system (\ref{redVME}) also inherits many of the properties of the VME, including K-conjugation$^{\rf{12},\rf{18}-\rf{23},\rf{1},\rf{3}}$ 
\vspace{-.0666in}
\alpheqn
\(
\tilde{T}(z)=T_{\gh}(z)-T(z) \sp\tilde{T}(z)T(w)=\reg
\)
\(
\tilde{L}_{J-L}^{ab}=(L_{J-L}^{ab})_{\gh}-L_{J-L}^{ab}\sp
\tilde{c}=c_{\gh}-c
\)
\(
(L_{J-L}^{ab})_{\gh} = \delta_{JL} \frac{\eta^{ab}}{2k+Q_{\sgb}} \sp c_{\gh}=\frac{2\l k \dg}{2k+Q_{\sgb}}
\)
\reseteqn
\vspace{.05in}
where $T_{\gh}$ is the affine-Sugawara construction$^{\rf{12},\rf{18},\rf{19}-\rf{21}}$ on $\gh=\oplus_{I=0}^{\l-1}\gb^I$, $\gb^I\cong \gb$.

It should be emphasized that many other solutions of the reduced VME in (\ref{redVME}) are known, including those coset constructions$^{\rf{12},\rf{18},\rf{22}}$ which are $\zl$-invariant on semisimple $\gh$ (see App.~E), as well as the unitary irrational solutions called (simply-laced $\gb$)$^q$ in Ref.~\rf{27}.
\subsection{Eigenvalue problem} \label{epsec}
In what follows we will need the solution of the eigenvalue problem
\vspace{.1in}
\namegroup{eprobgrp}
\alpheqn
\(
\sum_L \ws_{JL} U^\dagger(\s)_{L;r j}=U^\dagger(\s)_{J+\s;rj}=U^\dagger(\s)_{J;rj}\ E_r(\s) \label{eprob}
\)
\(
U^\dagger(\s)_{J+\l;rj}=U^\dagger(\s)_{J;rj}\sp
U^\dagger(\s)_{J;r+\r(\s),j}=U^\dagger(\s)_{J;r,j}
\)
\(
r,s=0,...,\r(\s)-1\hsp{1}\textup{(spectral index)}
\)
\(
j,l=0,...,\lors-1\hsp{1}\textup{(degeneracy index)}
\)         
\(
\s=0,...,\l-1
\)
\reseteqn
for each element $\omega$ of the automorphism group $\z_\l\subset Aut(g)$.  Because each $\omega$ is an orthogonal matrix, the matrix of eigenvectors can be taken to be 
unitary:
\(
\sum_JU_{rj;J}\ud_{J;sl}=\delta_{r,s\mod\r}\ \delta_{j,l \mod\lor}\sp
\sum_{r,j}\ud_{J;rj} \ U_{rj;L}=\delta_{J,L\mod\l}
\)
and we often omit the label $\sigma$ as seen here.  The spectral resolution of each automorphism 
\(
\w_{JL}=\sum_{r,j} \ud_{J;rj} \ E_r \ U_{rj;L} \label{specres}
\)
will also be useful below.

Solution of the eigenvalue problem is straightforward.  Because $\omega(h_\sigma)$ has order $\rho (\sigma)$, we know that its eigenvalues are the $\rho$th roots of unity, and we may choose the uniform ordering convention
\(
E_r(\s)=e^{-\frac{\tp  r}{\r(\sigma)}} \label{evconv}
\)
for the eigenvalues.  With this convention we may also choose
\namegroup{Ustar}
\alpheqn
\(
U^*_{r,j;J}=U_{-r,j;J}
\)
\(
\sum_J U_{r,j;J} U_{s,l;J}=\delta_{r+s,0 \mod\r} \ \delta_{j,l \mod\lor}
\sp
\sum_{r,j} U_{-r,j;J}U_{r,j;L}=\delta_{J,L\mod\l}
\)
\reseteqn
because each $\omega$ is real.

An explicit solution which satisfies all these conditions (see (\ref{Jper})) is
\(
U(\s)_{rj;J}=\isrs e^{-\frac{\tp N(\s)r(j-J)}{\l}}\delta_{j,J\mod\lors}\sp\usd_{J;rj}=\isrs e^{\frac{\tp N(\s)r(j-J)}{\l}}\delta_{j,J\mod\lors} 
\)
and this solution also satisfies the relation 
\(
U(\s)_{r,j;J}U(\s)_{s,l;J}=\delta_{j,l\mod\lors}U(\s)_{r+s,l;J} \label{John}
\)
(see (\ref{sumrule})).  The general solution $U$ of the eigenvalue problem (\ref{eprobgrp}) is given in App.~C, where it is also shown that the extra structure in $U$ corresponds to automorphisms of the orbifold affine algebra.
\subsection{The eigencurrents of sector $\s$}
We return to the currents $J_{aI}, \ I=0,...,\l-1$ on semisimple $\gh=\oplus_{I=0}^{\l-1}\gb^I$, $\gb^I\cong\gb$ which satisfy the affine algebra
\namegroup{ecaa}
\alpheqn
\(
J_{aJ}(z)J_{bL}(w)=\delta_{JL}[\frac{k\eta_{ab}}{(z-w)^2}+\frac{if_{ab}^{\ \ c}J_{cJ}(w)}{(z-w)}]+\reg \label{Jalg}
\)
\(
J_{aI}(ze^{\tp })=J_{aI}(z)\sp
\arange\sp
I,J,L=0,...,\l-1
\)
\reseteqn
and we recall the off-diagonal action
\(
J_{aJ}(z)\p=\ws_{JL}J_{aL}(z)\sp\ws\in\z_\l \label{ndJa}
\)
of the $\z_\lambda$ automorphisms on these currents.  Next, we define the \textit{eigencurrents} $\j$ of sector $\s$
\alpheqn
\(
\j_{a,r j}(z) \equiv \sqrt{\r(\s)} \sum_{J=0}^{\l-1}U(\s)_{rj;J}J_{a,J}(z)=\sum_{s=0}^{\r(\s)-1}e^{\frac{\tp N(\s)rs}{\r(\s)}} J_{a,\lors s+j}(z) \label{jJdef}
\)
\(
\j_{a;r+\r(\s),j}(z)=\j_{a;r,j}(z)
\)
\(
J_{aI}(z)=\isrs \sum_{r,j}\usd_{I;rj}\j_{a,rj}(z)
=\frac{1}{\r(\s)}\sum_{r=0}^{\r(\s)-1}e^{-\frac{\tp N(\s)r\lfloor\frac{\r(\s)}{\l} I\rfloor}{\r(\s)}}\j_{a,r,I\mod\lors}(z) \label{ec}
\)
\reseteqn
on which the automorphism (\ref{ndJa}) acts diagonally
\(
\j_{a,rj}(z)\p=\sqrt{\r(\s)} \sum_{I=0}^{\l-1}U(\s)_{rj;I} J_{aI}(z)\p =E_r(\s)\j_{a,rj}(z) \label{diagact}
\)
according to (\ref{ndJa}) and (\ref{specres}).  

Using (\ref{ecaa}), (\ref{jJdef}), (\ref{Ustar}) and (\ref{John}), we find that the eigencurrents satisfy the algebra
\alpheqn
\(
\j_{a,rj}(z)\j_{b,sl}(w)=\delta_{j l}[\frac{\r(\sigma) k\eta_{ab}
\delta_{r+s,0\mod\r(\sigma)}}{(z-w)^2}+\frac{if_{ab}^{\ \ c}\j_{c;r+s,j}(w)}{(z-w)}]+\reg \label{sseca}
\)
\(
\j_{a,rj}(ze^{\tp })=\j_{a,rj}(z)
\)
\(
r,s=0,...,\r(\sigma)-1\sp j,l=0,...,\frac{\l}{\r(\sigma)}-1
\)
\reseteqn
for each $\sigma = 0,...,\lambda-1$.

\subsection{The twisted currents of sector $\sigma$}
In abelian orbifold theory$^{\rf{28}-\rf{17}, \rf{9}, \rf{41}, \rf{42}}$, there is an orbifold sector for each element $\sigma = 0,...,\lambda -1$ of the automorphism group, where the trivial element $\sigma=0$ corresponds to the untwisted sector and $\sigma = 1,..,\lambda -1$ are the twisted sectors. In each sector
$\sigma$, one has the \textit{twisted currents} $\hat{J}$ which are in one to one correspondence with the eigencurrents $\j$ of that sector
\(
\j_{a,rj}(z)\sa\jh_{aj}^{(r)}(z)\equiv\jh_{a(j)}^{(r)}(z) \sp \s=0,...,\l-1. \label{tccorr}
\)
The aspects of this correspondence are twofold: In the first place, the twisted currents of sector $\sigma$ have the diagonal monodromy 
\(
\jh_{aj}^{(r)}(ze^\tp)=E_r(\s)\hsp{.03}\jh_{aj}^{(r)}(z)\sp E_r(\s)=e^{-\frac{\tp r}{\r(\s)}} \label{mono}
\)
which corresponds to the diagonal action (\ref{diagact}) of $h_\s$ on the eigencurrents.  Second, the twisted currents of sector $\sigma $ satisfy a twisted current algebra whose OPE form (but not its mode form) is isomorphic to the OPE's (\ref{sseca}) of the eigencurrents of that sector. 

Combining these two properties, we find that the twisted currents $\hat{J}$ of sector $\sigma$ satisfy
\namegroup{twistalg}
\alpheqn
\(
\jh_{aj}^{(r)}(z)\jh_{bl}^{(s)}(w) \hspace{-.03in} = \hspace{-.03in} \delta_{j l}[\frac{\r(\sigma) k\eta_{ab}\delta_{r+s,0\mod\r(\sigma)}}{(z-w)^2}+\frac{if_{ab}^{\ \ c}\jh_{cj}^{(r+s)}(w)}{(z-w)}]+\reg \label{TCOPE}
\)
\(
\hat{J}_{aj}^{(r)}(ze^{\tp}) = e^{-\frac{\tp r}{\r(\sigma)}} \hat{J}_{aj}^{(r)}
(z)
\)
\(
\hat{J}_{aj}^{(r)}(z) = \sum_{m \in \sz} \hat{J}_{aj}^{(r)}(m+\frac{r}{\r(\sigma)}) z^{-1-m-\frac{r}{\r(\sigma)}}
\)
\(
\arange\sp
r,s=0,...,\r(\sigma)-1\sp
j,l=0,...,\frac{\l}{\r(\sigma)}-1.
\)
\reseteqn
The twisted current algebra (\ref{TCOPE}) is recognized\footnote{The preceding derivation of orbifold affine algebra was indicated but not completed in Ref.~\rf{9}.} as the orbifold affine algebra
\(
\gb_{\r(\s)}\equiv\oplus_{j=0}^{\frac{\l}{\r(\s)}-1} \gb_{\r(\s)}^j \label{grho}
\)
on the semisimple Lie algebra $\oplus_{j=0}^{\lors-1}\gb^j$, where each of the $\lors$ simple copies $\gb^j$ is taken at orbifold affine order $\rho(\sigma)$. We emphasize that the \textit{order} $\r(\s)$ of the orbifold affine algebra $\gb_{\r(\s)}$ is also the \textit{order} of the automorphism $h_\s$.

The untwisted sector has $\r(0) =1$ and hence $r,s=0$, so the orbifold algebra (\ref{twistalg}) reduces in this case to the untwisted affine Lie algebra (\ref{Jalg}) on semisimple $\gh$:
\(
\gh=\oplus_{j=0}^{\l-1}\gb^j=\oplus_{J=0}^{\l-1}\gb^J 
\sp J_{aJ}(z) \equiv \hat{J}_{aJ}^{(0)}(z).
\)
When an index $j$ has range $0$ to $\l-1$, it is our convention to replace that index by a capital letter as a reminder that we have returned to the semisimplicity of the VME in (\ref{cycdefs}).

The representation theory of orbifold affine algebra was obtained
via the \textit{orbifold induction procedure} in Ref.~\rf{9}. In particular,
we know that
\alpheqn
\(
\hat{J}^{(r)}_{aj} (m + \frac{r}{\r(\s)}) |0\rangle_\sigma =0 \ \ \textup{when} \ (m+\frac{r}{\r(\s)}) \geq 0
\)
\( r=0,...,\r(\s)-1\sp j= 0,...,\frac{\lambda}{\rho(\sigma)} -1
\)
\reseteqn
where $|0\rangle_\sigma$ is the ground state of sector $\sigma$. 

We can also construct twisted currents $\hat{\j}$ with mixed monodromy in each sector
\alpheqn
\(
\hat{\j}_{aJ}(z)\equiv \frac{1}{\sqrt{\r(\s)}} \sum_{r,j} \usd_{J;rj} \ \hat{J}_{aj}^{(r)}(z)\sp
\hat{\j}_{aJ}(ze^{\tp})=\ws_{JL} \ \hat{\j}_{aL}(z) \label{Jj}
\)
\(
\hat{\j}_{aJ}(z)\hat{\j}_{bL}(w)=\delta_{JL}[\frac{k\eta_{ab}}{(z-w)^2}+\frac{if_{ab}^{\ \ c}\hat{\j}_{cJ}(w)}{(z-w)}]+\reg \label{Jhatalg}
\)
\reseteqn
whose OPE's in (\ref{Jhatalg}) are isomorphic to the original untwisted current algebra in (\ref{Jalg}).  The inverse of (\ref{Jj}),
\(
\hat{J}_{aj}^{(r)}(z) =\sqrt{\r(\s)} \ \sum_{J=0}^{\l-1} U(\s)_{rj;J} \ \hat{\j}_{aJ}(z) \label{jJ}
\) 
gives the twisted currents $\hat{J}$ as the monodromy decomposition of $\hat{\j}$.  

\subsection{The stress tensor of sector $\sigma$} \label{Tofsig}
We begin with the stress tensor $T$ of any $\z_\lambda$(permutation)-invariant CFT (see
Subsec.~\ref{CVMEsec}) and rewrite it, using (\ref{jJdef}), in terms of the eigencurrents $\j$ of each sector
\alpheqn
\(
T(z) = \ \sum_{J,L}L_{J-L}^{ab}:J_{aJ}(z)J_{bL}(z):\ \ = \ \sum_{r,s}\sum_{j,l}\lr_{r,s}^{a(j)b(l)}(\s):\j_{a,rj}(z)\j_{b,sl}(z): \label{TJJjj}
\)
\(
\lr_{r,s}^{a(j)b(l)}(\s)\equiv\frac{1}{\r(\s)}\sum_{J,L}L_{J-L}^{ab}\usd_{J;rj}\usd_{L;sl}\ . \label{luu}
\)
\reseteqn
Then we may use the OPE isomorphism (\ref{tccorr}) to deduce that the stress tensor $\hat{T}_\s$ of sector $\sigma$ is
\alpheqn
\(
T(z)\ \sa\ \hat{T}_\s (z) \label{Tarrow}
\)
\(
\hat{T}_\s (z)=\sum_{r,s}\sum_{j,l}\lr_{r,s}^{a(j)b(l)}(\s):\jh_{aj}^{(r)}(z)
\jh_{bl}^{(s)} (z):.
\)
\reseteqn      
The Virasoro property of $\hat{T}_\s$ follows from that of $T$ because the OPE's of $\hat{J}$ and $\j$ are isomorphic.

Using Eq. (\ref{halftrick}), one finds that the coefficients $\lr$ have the property
\alpheqn
\(
\lr_{r,s}^{a(j)b(l)}(\s)=\delta_{r+s,0\mod\r(\s)}\lr_{r}^{a(j)b(l)}(\s) 
\label{lrnomono}
\)
\(
\hat{T}_\s (ze^\tp)=\hat{T}_\s (z) \label{tmono}
\)
\reseteqn
which guarantees the trivial monodromy shown for $\hat{T}_\s$ in (\ref{tmono}).  Furthermore using (\ref{trick}) we can evaluate\footnote{An alternate form of $\lr_r$ is given in (\ref{loldL}).} $\lr_r$ in (\ref{lrnomono}) to obtain the explicit form of the stress tensor of sector $\s$ 
\namegroup{mainres}
\alpheqn
\(
\hat{T}_\s (z)=\sum_{r=0}^{\r(\sigma)-1} \ \sum_{j,l=0}^{\frac{\l}{\r(\sigma)}-1} \lr_r^{a(j)b(l)}(\s):\jh_{aj}^{(r)}(z)\jh_{bl}^{(-r)}(z): \sp \s=0,...,\l-1 \label{twstress}
\)
\getletter{themaplett}
\(
\lr_{r}^{a(j)b(l)}(\s)=\frac{1}{\r(\sigma)}\sum_{s=0}^{\r(\sigma)-1} e^{-\frac{\tp N(\s)rs}{\r(\sigma)}} L^{ab}_{\frac{\l}{\r(\sigma)} s+j-l} \label{themap}
\)
\(
L^{ab}_{\frac{\l}{\r(\sigma)} s+j-l}=\sum_{r=0}^{\r(\sigma)-1} e^{\frac{\tp N(\s)rs}{\r(\sigma)}} \lr_{r}^{a(j)b(l)}(\s) \label{theinverse}
\)
\
\getletter{allsymslett}
\(
L^{ab}_K=L^{ba}_{-K}=L^{ab}_{K\pm\l}\sp K=0,...,\l-1
\label{allsyms}
\)
\(
\arange\sp
r,s=0,...,\r(\sigma)-1\sp 
j,l=0,...,\frac{\l}{\r(\sigma)}-1.
\)
\reseteqn
The explicit form of this set of duality transformations is one of the central results of the paper.

In particular, \textbf{the duality transformations (\ref{themap}) give the inverse inertia tensors $\mathbf{\lr}$ of all the twisted sectors of each orbifold}
\(
\az
\sp \z_\l \subset Aut(g) \sp g=\oplus_{I=0}^{\l-1} \gb^I \sp \gb^I \cong \gb
\)
\textbf{where} $\mathbf{A(\z_\lambda)}$ \textbf{is the} $\z_\lambda$\textbf{(permutation)-invariant CFT on $g$ whose inverse inertia tensor is} $L$.  

In fact, the stress tensor $T$ in (\ref{TJJjj}) of the untwisted sector can also be obtained from $\hat{T}_\s$ in (\ref{mainres}) by choosing $\s=0$.  This gives the connection
\alpheqn
\(
L_{J-L}^{ab} \equiv \lr_0^{a(J) b(L)}(\s=0) \sp T(z) \equiv \hat{T}_{\s=0}(z)
\)
\(
r=s=0 \sp J,L=0,...,\l-1
\)
\reseteqn
because $\r(0)=1$.

The inverse inertia tensors $\lr$ of the twisted sectors inherit the symmetries
\namegroup{inhersym}
\alpheqn
\ba
\lr_r^{a(j)b(l)}(\s) &=& \lr_{-r}^{b(l) a(j)}(\s) \comment{$(a(\cdot),r\lra b(\cdot),-r)$} \label{actualsymm}\\
&=& \lr_{r \pm \r(\s)}^{a(j)b(l)}(\s) \comment{(periodicity)}
\ea
\ba
\left.
\begin{array}{ll}
\lr_r^{a(j+\lors)b(l)}(\s)=e^{\frac{\tp Nr}{\r(\s)}} \lr_r^{a(j)b(l)}(\s) \\
\lr_r^{a(j)b(l+\lors)}(\s)=e^{-\frac{\tp N r}{\r(\s)}} \lr_r^{a(j)b(l)}(\s)
\end{array} \right\} \comment{(quasiperiodicity)}
\ea
\reseteqn
from (\ref{themap}) and the ($a(\cdot) \lra b(\cdot)$) symmetry of the untwisted sector $L$ in (\ref{zlsyms}).

The twisted sectors also inherit a residual $\z_{\lambda/\r} \subset \z_\lambda$ symmetry
\namegroup{inherresid}
\alpheqn
\(
\lr_r^{a(j+1)b(l+1)}(\s)=\lr_r^{a(j)b(l)}(\s) \comment{(residual $\z_\frac{\l}{\r(\sigma)}$ invariance)} \label{inhresid}
\)
\(
j,l=0,...,\frac{\l}{\r(\sigma)}-1
\)
\reseteqn
from the $\z_\lambda$ symmetry of the untwisted sector with $\r(0)=1$. In
orbifold theory, one must further mod out by this residual symmetry
in each sector $\sigma$.

In App.~D, we note that unitarity of the untwisted sector of any orbifold \taz \ implies unitarity of all the twisted sectors.   Moreover, App.~E collects some simple examples of the duality transformations in (\ref{mainres}).
\subsection{Central charge and conformal weights} \label{charge&weight}
Because of the correspondences in (\ref{tccorr}) and (\ref{Tarrow})
\[ \j(z) \sa \hat{J}(z) \sp T(z) \sa \hat{T}_\s (z)
\]
we also know that the central charge $\hat{c}(\s)$ of $\hat{T}_\s$ is the same in all sectors 
\namegroup{allsectc}
\alpheqn
\(
\hat{c}(\s)=c = 2\l k \eta_{ab}L^{ab}_0
=2 \sum_{j,l=0}^{\frac{\l}{\r(\s)}-1} \hat{G}_{a(j)b(l)}(\s) \sum_{r=0}^{\r-1} \lr_r^{a(j) b(l)}(\s)  \nonumber
\)
\(
\hat{G}_{a(j)b(l)}(\s) \equiv \r(\s)k\eta_{ab}\delta_{jl}\sp  \s=0,...,\l-1
\)
\reseteqn
where we have used (\ref{cLzero}), and (\ref{lzero}).  The last form agrees with the central charge given in (\ref{delta}), using $\gb_{\r(\s)}$ defined in (\ref{grho}).

Following Ref.~\rf{5}, the conformal weight of the ground state $|0\rangle_\sigma$ of each sector is easily computible from each $\hat{T}_\s$.  One uses the identities
\alpheqn
\(
 \hat{T}_\s (z)=\sum_{m \in \sz} L_\s (m) z^{-m-2}, \hspace{.3in} L_\s (m \geq 0) |0\rangle_\sigma = \delta_{m,0}\hat{\Delta}_0(\sigma)|0\rangle_\sigma
\)
\( 
\hat{J}_{aj}^{(r)}(m+\frac{r}{\r(\s)}) |0\rangle_\sigma = 0 \ \textup{when} \ (m+\frac{r}{\r(\s)}) \geq 0  \label{deleteme}
\)
\(
:\jh_{aj}^{(r)}\jh_{bl}^{(-r)}:(m=0)|0\rangle_\sigma=\hat{G}_{a(j)b(l)}(\s)\frac{r(\r(\s)-r)}{2\r^2(\s)}|0\rangle_\sigma \sp r=0,...,\r(\s)-1
\)
\reseteqn
along with the form of $\hat{T}_\s$ in (\ref{twstress}). The result is
\vspace{-.0666in} 
\namegroup{OVMEcw}
\alpheqn
\ba
\hat{\Delta}_0(\s)&=&\sum_{j=0}^{\lors-1}\r(\s) k \eta_{ab}\sum_{r=0}^{\r(\s)-1}
\lr_r^{a(j) b(j)} (\sigma) \ \frac{r(\r(\s)-r)}{2\r^2(\s)} \label{cwlr} \\
&=&\left\{
\begin{array}{cl}
0&,\ \s=0\\
\frac{\l k\eta_{ab}}{4\r^2(\s)}(\frac{\r^2(\s)-1}{3}L_0^{ab}-{\displaystyle{\sum_{s=1}^{\r(\s)-1}}} csc^2(\frac{\pi N(\s)s}{\r(\s)})\hsp{.02} L^{ab}_{\frac{\l}{\r(\s)}s})&,\ \s=1,...,\l-1 \label{cwLsummed}
\end{array}
\right.
\ea 
\reseteqn
where we have used the duality transformations (\ref{themap}) to express the conformal weights in terms of the inverse inertia tensor $L$ of the untwisted sector of each orbifold.  

\subsection{The OVME and $A(\subsecd_{\hsp{.04}\l})/\subsecz_\l$} \label{OVMEsec}
In Sec.~\ref{Jansec}, we showed that the OVME in (\ref{OVMEgrp}) is a collection of twisted sectors of the permutation orbifolds
\(
\ad \label{ad} 
\)
but we obtained there only one twisted sector of each \tad.

On the other hand, we have now constructed all the twisted sectors of all the permutation orbifolds
\(
\az\supset\ad
\)
and, correspondingly, the set of duality transformations (\ref{mainres}) is easily specialized to the higher symmetry of $A(\d_\l)$ in (\ref{DVME}).  This gives the duality transformations for all the sectors of all the orbifolds \tad
\namegroup{everysec}
\alpheqn
\(
\hat{T}_\s (z)=\sum_{r=0}^{\r(\sigma)-1} \ \sum_{j,l=0}^{\frac{\l}{\r(\sigma)}-1} \lr_r^{a(j)b(l)}(\s):\jh_{aj}^{(r)}(z)\jh_{bl}^{(-r)}(z): \sp \s=0,...,\l-1 \label{twstress2}
\)
\(
\lr_{r}^{a(j)b(l)}(\s)=\frac{1}{\r(\sigma)}\sum_{s=0}^{\r(\sigma)-1} e^{-\frac{\tp N(\s)rs}{\r(\sigma)}} L^{ab}_{\frac{\l}{\r(\sigma)} s+j-l}\ 
\)
\(
L_{J-L}^{ab}=L_{J-L}^{ba}=L_{L-J}^{ab}=L_{J-L \pm \l}^{ab}
\)
\reseteqn
where $L$ is the inverse inertia tensor of each $\d_\l$-invariant CFT.  The formulae for $\hat{c}$ and $\hat{\Delta}_0$ in Subsec.~\ref{charge&weight} hold for these orbifolds as well.

The special case $\s=1$ of the result (\ref{everysec}) is the first duality transformation of Subsec.~\ref{fdt}.  To see this, note that $\r(1)=\l$, $M(1)=N(1)=1$ and $j=l=0$, so that
\(
\lr_{r}^{ab}\equiv\lr_{r}^{a(0)b(0)}(\s=1) =\frac{1}{\l}\sum_{K=0}^{\l-1} e^{-\frac{\tp Kr}{\l}} L^{ab}_{K}=\lr_{-r}^{ab} \label{lcase}
\)
is obtained for $\s=1$.  This is just the $S=1$ case of our first duality transformation (\ref{Janinv}) given explicitly in (\ref{simpcase}).

The symmetry $\lr_r^{ab}=\lr_{-r}^{ab}$ in (\ref{lcase}) confirms that the sector $\s=1$ with $\r(1)=\l$ is in the OVME (see Eq. (\ref{combo})).  Similarly all sectors with $\r(\s)=\l$
\(
\lr_r^{a(0)b(0)}(\r(\s)=\l)=\frac{1}{\l}\sum_{K=0}^{\l-1}e^{-\frac{\tp N(\s)Kr}{\l}}L_K^{ab}=\lr_{-r}^{a(0)b(0)}(\r(\s)=\l)
\)
are in the OVME, and one may ask whether the sectors with $\r(\s)\neq\l$ are solutions of the OVME.  Unfortunately the answer is in general no because the generic sector (\ref{everysec}) of \tad\ does \textit{not} satisfy the symmetry (\ref{combo}),
\(
\lr_r^{a(j)b(l)}(\s) \neq \lr_{-r}^{a(j)b(l)}(\s) 
\)
which is required for the sector to be in the OVME.

On the other hand, the OVME contains all the sectors of many simple but important examples of \tad\ orbifolds, including:

\vspace{.2in}
\noindent
\bu{ the cyclic copy orbifolds} $(\times_{I=0}^{\l-1}A_I)/\z_\l$\\
\bu{ the cyclic orbifolds on $J^{\diag(\s)}$}\\
\bu{ interacting coset orbifolds}\\
\bu{ \tad\ with prime $\l$ \\
\bu{ $A(\dl)/\z_\l$ for $\l \leq 7$

\vspace{.2in}
\noindent
These and related examples are discussed explicitly in App.~E.

\newsection{The Cyclic OVME}
\label{covmesec}
In this section, we use the duality transformations to  construct an extended or cyclic OVME which contains the stress tensors of all the sectors of all the permutation orbifolds
\(
\az\supset\ad
\)
and which contains the OVME as a consistent subansatz
\(
\textup{Cyclic OVME $\supset$ OVME}.
\)
We note in particular that the duality transformations are sufficient to derive the cyclic OVME from the VME without having to do any additional OPE's explicitly.

\subsection{Return to general semisimplicity}
Our starting point in this derivation is a return to the \textit{general semisimplicity}
\(
\gh=\oplus_{I=0}^{\l-1}\gb^I \sp \gb^I= \oplus_{\alpha=0}^{S-1} \gb^{\a I} \sp \gb^I \cong \gb
\)
studied for the VME in Sec.~\ref{Jansec}.  Here $\a$ runs over generically distinct simple algebras and $I$ runs over copies of the algebra $\gb$.  (Recall that we chose the case $S=1$ and hence $\a,\b=0$ for simplicity in Sec.~\ref{mainsec}.)

On this semisimplicity, the general affine-Virasoro construction takes the form
\alpheqn
\(
T(z)=L^{a(\alpha J)b(\beta L)}:J_{a(\alpha J)}(z) J_{b(\beta L)}(z):
\)
\(
L^{a(\alpha J) b(\beta L)} = L^{b(\beta L) a(\alpha J)}\comment{($a(\cdot)\lra b(\cdot)$\ symmetry)}
\)
\(
J,L=0,...,\l-1\sp\alpha,\beta=0,...,S-1
\)
\reseteqn
and we define the $\z_\l$(permutation)-invariant CFT's by the relations
\alpheqn
\(
J_{a(\alpha J)}(z)\p=\ws_{JL} \ J_{a(\alpha L)}(z)
\)
\(
L^{a(\alpha J\p) b(\beta L\p)}\ws_{J\p J} \ \ws_{L\p L} =L^{a(\alpha J)b(\beta L)}\sp \forall \ h_\s\in\z_\l\subset Aut(\gh). \label{Lauto}
\)
\reseteqn
The $\z_\l$ automorphisms $\w$ are the same as in (\ref{wdef}) and so the eigenvalue problem (\ref{eprob}) and its solution are unchanged.

The solutions to (\ref{Lauto}) are
\alpheqn
\ba
L^{a(\alpha, J+1) b(\beta, L+1)}=L^{a(\alpha, J) b(\beta, L)}&=&L_{J-L}^{a(\alpha) b(\beta)} \comment{($\zl$ invariance)}\\
&=&L_{L-J}^{b(\beta)a(\alpha)}\comment{($a(\cdot)\lra b(\cdot)$\ symmetry)}
\ea
\reseteqn
and the reduced VME of the $\z_\l$(permutation)-invariant CFT's is 
\ba
L_{J-L}^{a(\alpha)b(\beta)}&=&2\sum_{M=0}^{\l-1}L_{J-L-M}^{a(\alpha)c(\gamma)} G_{c(\gamma)d(\delta)} L_M^{d(\delta)\b(\beta)} + L_{J-L}^{c(\gamma)d(\delta)} L_{J-L}^{e(\epsilon) f(\phi)} f_{c(\gamma) e(\epsilon)}^{\ \ \ \ \ \ \ a(\alpha)} f_{d(\delta) f(\phi)}^{\ \ \ \ \ \ \ b(\beta)} \nonumber \\
&\ & + L_0^{c(\gamma) d(\delta)} f_{c(\gamma) e(\epsilon)}^{\ \ \ \ \ \ \ f(\phi)}[f_{d(\delta) f(\phi)}^{\ \ \ \ \ \ \ a(\alpha)} L_{L-J}^{b(\beta) e(\epsilon)}+f_{d(\delta) f(\phi)}^{\ \ \ \ \ \ \ b(\beta)} L_{J-L}^{a(\alpha) e(\epsilon)}] \label{SSCVME}
\ea
where summation over pairs of repeated Greek indices is implicit.  The metric and structure constants here are defined in (\ref{OVMEgf}).  

This leads us to the eigencurrents $\j$ of each sector
\alpheqn
\(
\j_{a \alpha,rj}(z)\equiv\sqrt{\r}\sum_{J=0}^{\l-1} U_{rj;J} J_{a (\alpha J)}(z)\sp J_{a (\alpha J)}(z)=\isr\sum_{r,j} \ \ud_{J;rj}\j_{a \alpha,rj}(z)
\)
\(
\j_{a \alpha,rj}(z)\p=E_r \ \j_{a \alpha,rj}(z)\sp j,l=0,...,\lor-1
\)
\reseteqn
and then to the duality transformations
\namegroup{dtlist}
\alpheqn
\(
\hat{T}_\s (z)=\sum_{r,j,l} \lr_r^{a(\alpha j) b(\beta l)}(\s) :\hat{J}_{a (\alpha j)}^{(r)}(z) \hat{J}_{b(\beta l)}^{(-r)}(z):
\)
\getletter{SSmapslett}
\(
\lr_r^{a(\alpha j) b(\beta l)}(\s) =\frac{1}{\r(\s)} \sum_{s=0}^{\r-1} e^{-\frac{\tp N rs}{\r(\s)}} L_{\lors s+j-l}^{a(\alpha) b(\beta)}  \sp
L_{\lors s+j-l}^{a(\alpha) b(\beta)}=\sum_{r=0}^{\r-1} e^{\frac{\tp Nrs}{\r(\s)}}\lr_r^{a(\alpha j) b(\beta l)}(\s) \label{SSmaps}
\)
\getletter{SSLsymmlett}
\(
L_{\lor s+j-l}^{a(\alpha) b(\beta)}=L_{-\lor s+l-j}^{b(\beta) a(\alpha)}=L_{\lor s+j-l \pm \l}^{a(\alpha) b(\beta)} 
\)
\ba
\lr_r^{a(\alpha j) b(\beta l)}(\s) &=& \lr_{-r}^{b(\beta l) a(\alpha j)}(\s)\comment{($a(\cdot),r\lra b(\cdot),-r$ symmetry)} \\
&=&\lr_{r\pm\r(\s)}^{a(\alpha j) b(\beta l)}(\s)\comment{(periodicity)}\\
\lr_r^{a(\alpha, j+1) b(\beta, l+1)}(\s) &=& \lr_r^{a(\alpha j) b(\beta l)}(\s)\comment{(residual $\z_\frac{\l}{\r(\s)}$ invariance)}  \label{resid}
\ea
\reseteqn
which generalize the duality transformations of Sec.~\ref{mainsec}.  As a check on these results, we may reconsider the less general semisimplicity $S=1$ of Sec.~\ref{mainsec}: With the definition
\(
a(0j)\equiv a(j)\sp b(0j)\equiv b(j)
\)
one sees that the results (\ref{dtlist}) reduce to Eqs. (\ref{mainres}), (\ref{inhersym}) and (\ref{inherresid}).

\subsection{From the $\subsecz_\l$-invariant CFT's to the cyclic OVME}
For this derivation, we first consider any sector $\s$ for which $\r(\s)=\l$.  In any such case we have $j=l=0$, and the duality transformations (\ref{dtlist} \ref{SSmapslett},\ref{SSLsymmlett}) become
\namegroup{ssduality}
\alpheqn
\(
\lr_r^{a(\alpha) b(\beta)} \equiv \lr_r^{a(\alpha 0) b(\beta 0)} 
\)
\(
\lr_r^{a(\alpha) b(\beta)} = \frac{1}{\l} \sum_{K=0}^{\l-1} e^{-\frac{\tp N rK}{\l}} L_{K}^{a(\alpha) b(\beta)} \sp
L_{K}^{a(\alpha) b(\beta)} = \sum_{r=0}^{\l-1} e^{\frac{\tp N rK}{\l}} \lr_r^{a(\alpha) b(\beta)}.
\)
\reseteqn
Next, for each $\z_\l$(permutation)-invariant CFT (i.e. $L$ satisfies (\ref{SSCVME})) we may reverse the logic of Sec.~\ref{Jansec} to find the equation satisfied by $\lr$.  The result is 
\ba
\lr_r^{a(\alpha) b(\beta)} &=&  2 \lr_r^{a(\alpha) c(\gamma)}  \hat{G}_{c(\gamma)d(\delta)} \lr_r^{d(\delta)b(\beta)} \nonumber \\
&\ & - \sum^{\lambda-1}_{s=0} \lr_s^{c(\gamma) d(\delta)} [\lr_{r-s}^{e(\epsilon) f(\phi)} f_{c(\gamma) e(\epsilon)}^{\ \ \ \ \ \ \ a(\alpha)}f_{d(\delta) f(\phi)}^{\ \ \ \ \ \ \ b(\beta)} + f_{c(\gamma) e(\epsilon)}^{\ \ \ \ \ \ \ f(\phi)}f_{d(\delta) f(\phi)}^{\ \ \ \ \ \ \ a(\alpha)} \lr_{-r}^{b(\beta) e(\epsilon)} \nonumber \\
&\ &+f_{c(\gamma) e(\epsilon)}^{\ \ \ \ \ \ \ f(\phi)}f_{d(\delta) f(\phi)}^{\ \ \ \ \ \ \ b(\beta)} \lr_{r}^{a(\alpha) e(\epsilon)}  ] \label{eOVME}
\ea
where the metric and structure constants are those defined in (\ref{OVMEgf}).

To simplify the form of this equation, we introduce the composite notation
\(
a(\alpha) \rightarrow a, \ b(\beta) \rightarrow b
\)
and summarize the new system as follows\footnote{The form of $\hat{c}$ and $\hat{\Delta}_0$ follow from (\ref{allsectc}) and (\ref{cwlr}).}
\namegroup{COVMEc}
\alpheqn
\(
\hat{T}(z)=\sum_{r=0}^{\l-1} \lr_r^{ab}:\hat{J}_a^{(r)}(z)\hat{J}_b^{(-r)}(z):
\)
\getletter{COVMElet}
\(
\lr_r^{ab} =  2 \lr_r^{ac}  \hat{G}_{cd} \lr_r^{db} - \sum^{\lambda-1}_{s=0} \lr_s^{cd} [\lr_{r-s}^{ef} f_{ce}^{\ \ a}f_{df}^{\ \ b} + f_{ce}^{\ \ f}f_{df}^{\ \ a} \lr_{-r}^{b e}+ f_{ce}^{\ \ f}f_{df}^{\ \ b} \lr_r^{a e}] \label{ceOVME}
\)
\getletter{perlet}
\(
\lr_{r \pm \l}^{ab}=\lr_r^{ab}\comment{(periodicity)}
\)
\getletter{ablet}
\(
\lr_r^{ab}=\lr_{-r}^{ba} \comment{$(a(\cdot),r \lra b(\cdot),-r)$}
\)
\(
\hat{c}=2\hat{G}_{ab}\sum_{r=0}^{\l-1} \lr_r^{ab}\sp\hat{\Delta}_0=\hat{G}_{ab}\sum_{r=0}^{\l-1} \lr_r^{ab}\frac{r(\l-r)}{2\l^2}
\)
\(
\gb=\oplus_{\a=0}^{S-1} \gb^\a \sp \hat{G}_{ab}=\stackrel{S-1}{\pa} \hspace{-.1in}  \hat{k}_\a \eta_{ab}^{\a} \sp \hat{k}_\a=\l k_\a 
\)
\(
a,b=1,...,\dg \sp  r,s=0,...,\l-1
\)
\reseteqn
where (\ref{COVMEc}\ \ref{COVMElet}-\ref{ablet}) is the \textit{cyclic orbifold Virasoro master equation} (cyclic OVME) at integer order $\l$.  The cyclic OVME, which also reduces to the VME at $\l=1$, is another central result of this paper\footnote{Although the system (\ref{COVMEc}) was derived here by duality from the VME, we have also checked these results by direct OPE computation as in Ref.~\rf{5}.}.

We have checked that the cyclic OVME in the form (\ref{COVMEc}) also contains the other sectors with $\r \neq \l$ in the duality transformations (\ref{SSmaps}).  To see this explicitly, one needs to consider the cyclic OVME with the explicit semisimplicity
\alpheqn
\(
a \rightarrow a(\alpha j), b \rightarrow b(\beta l)
\)
\(
\gb_{\r(\s)}=\oplus_{j=0}^{\lors-1}\gb_{\r(\s)}^j\sp
\gb^j\cong \gb\sp
\gb=\oplus_{\a=0}^{S-1}\gb^\a \label{covmeg}
\)
\reseteqn
discussed above and for the VME in Sec.~\ref{Jansec}.  In particular, the other sectors with $\r \neq \l$ are found at order $\r$ of the cyclic OVME with the residual $\z_{\l/\r}$ symmetry (\ref{resid}).  Solutions at order $\r$ with no residual $\z_{\l/\r}$ symmetry are sectors of $A(\z_\r)/\z_\r$ orbifolds.

One concludes that \textbf{the cyclic OVME contains all the sectors of all the orbifolds} $\mathbf{A(\z_\l)/\z_\l}$\textbf{, where} $\mathbf{A(\z_\l)}$ \textbf{is any} $\mathbf{\z_\l}$\textbf{(permutation)-invariant CFT.} 

It is instructive to compare the cyclic OVME in (\ref{COVMEc}) to the OVME in (\ref{OVMEgrp}), noting the small but important differences in their form and the symmetries required for each.  In particular, the symmetries of the solutions of the OVME imply the symmetries of the solutions of the cyclic OVME
\(
\lr_r^{ab}=\lr_{r\pm\l}^{ab}=\lr_{\l\pm r}^{ab}=\lr_r^{ba}\rightarrow
\lr_r^{ab}=\lr_{r\pm\l}^{ab}=\lr_{-r}^{ba} \label{symarrow}
\)
but not necessarily vice versa.  For this reason, the cyclic OVME (\ref{COVMEc}) reduces to the OVME (\ref{OVMEgrp}) when the ``gauge condition'' (\ref{gcond}) of the OVME is used in the cyclic OVME (and $s\rightarrow \l-s$ in the sums). 

It follows that the OVME is included as a consistent subansatz of the cyclic OVME, which is consistent with the fact that 
\(
\ad\subset\az.
\)
For the special case of $\l=2$, the arrow of (\ref{symarrow}) is reversible because $r=-r \ \textup{mod} \ 2$.  This tells us that the OVME and the cyclic OVME are identical at $\l=2$, which reflects the fact that $\d_2 \cong \z_2$.

\subsection{Technology of the cyclic OVME} \label{techno}

We have provided two computational approaches to the orbifolds $A(\z_\l)/\z_\l$.  In the more conventional approach shown in Fig.~1, 

\begin{picture}(350,140)(0,0)
\put(165,120){\line(1,0){120}}
\put(290,118){$\l$}
\put(165,85){\line(1,0){120}}
\put(290,83){$\r$}
\put(165,50){\line(1,0){120}}
\put(290,48){1}
\put(220,50){\vector(0,1){70}}
\put(220,50){\vector(1,1){35}}
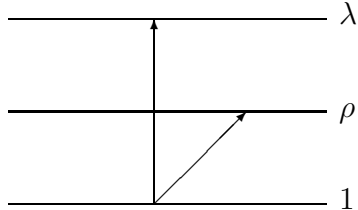
\put(155,20){Fig.~1 Starting from the VME}
\end{picture}

\noindent
one starts in the untwisted sector $L$ of any $\z_\l$(permutation)-invariant CFT and constructs all the twisted sectors $\lr$ via the duality transformations (\ref{themap}). This is the approach followed for simple examples in App.~E.  Alternately, one may start by finding the twisted sectors $\lr$ directly by solving the cyclic OVME.  Both approaches have their own merits and, indeed, ``new'' solutions have been found at both ends of the problem\footnote{The VME solutions$^{\rf{27}}$ known as (simply-laced $\gb$)$^q$ have an $S_q$(permutation)-invariance (when $\theta(J)=0$), and so these constructions are solutions of the reduced VME's in (\ref{DVME}) and (\ref{redVME}) when $q=\l$.  Used in (\ref{themap}), the solutions (simply-laced $\gb$)$^q$ give many examples of new $A(S_q)/\z_q\subset A(\z_q)/\z_q$ orbifolds.   Moreover, many new solutions of the OVME were found in Ref.~\rf{5}, including the Lie $\gb$-invariant constructions discussed below.}.  Consequently, we give here a brief discussion of the cyclic OVME in its own right.

The cyclic OVME (\ref{ceOVME}) is a set of 
\(
n_C(\gb,\l)=\dg\{\frac{\l(\dg-1)}{2}+\hl+1\} \label{nCOVME}
\)
coupled quadratic equations and unknowns, so the number of physically inequivalent solutions to the cyclic OVME expected at each level $\hat{k}$ is 
\(
N_C(\gb,\l)=2^{n_C({\sgb},\l)-\mbox{\scriptsize{\textup{dim}}}\sgb}.
\)  
These numbers $n_C$ and $N_C$ are the same numbers (\ref{nc}) and (\ref{NCeq}) found for the reduced VME of the $\z_\l$(permutation)-invariant CFT's. 

Many properties of the OVME persist in the cyclic OVME.  In particular K-conjugation 
\alpheqn
\(
\tilde{\hat{T}\hspace{.035in}}_{\hspace{-.05in}\s} (z)=\hat{T}_{\sgb_{\r(\s)}}(z)-\hat{T}_\s (z) \label{kconj}\sp
\tilde{\hat{T}\hspace{.035in}}_{\hspace{-.05in}\s} (z) \hat{T}_\s (w)=\reg\sp
\tilde{\hat{c}\hspace{.018in}}=\hat{c}_{\sgb_{\r(\s)}}-\hat{c} \label{allsecKconj}
\)
\(
\gb_{\r(\s)} = \oplus_{j=0}^{\frac{\l}{\r(\s)}-1} \gb_{\r(\s)}^j 
\)
\reseteqn
persists in every sector $\s$, with order $\r(\s)$.  Here $\gb_{\r(\s)}$ is the orbifold affine algebra (\ref{twistalg}) and $\hat{T}_{\sgb_{\r(\s)}}$ is the orbifold affine-Sugawara construction in sector $\sigma$ (see Ref.~\rf{5} and Eq.~(\ref{WZW})).

Moreover, the duality transformations in (\ref{SSmaps}) are invertible, so they can be used to construct all the sectors of any \taz\ orbifold from any particular solution of the cyclic OVME.  As a simple case, we consider the formula for the sectors $\lr(\s)$ at order $\r(\s)$ which are generated by a given solution $\lr_r^{ab}$ of the cyclic OVME on simple $g$ at order $\l\geq 2$, as shown in Fig.~2.

\begin{picture}(350,140)(0,0)
\put(165,120){\line(1,0){120}}
\put(290,118){$\l$}
\put(165,85){\line(1,0){120}}
\put(290,83){$\r$}
\put(165,50){\line(1,0){120}}
\put(290,48){1}
\put(220,120){\vector(0,-1){70}}
\put(220,120){\vector(1,-1){35}}
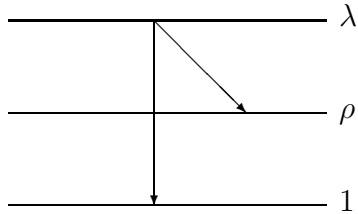
\put(140,20){Fig.~2 Starting from the cyclic OVME}
\end{picture}

\noindent
To proceed, we assign sector $\s_0$, with $\r(\s_0)=\l$, to the solution $\lr_r^{ab}$ where $\s_0$ can be chosen to be any integer from $1$ through $\l-1$, so long as $\s_0$ and $\l$ are relatively prime.  We also need the integers $N(\s_0), N(\s)$ and $P(\s,\s_0)$:
\alpheqn
\(
N(\s_0) \hspace{.03in}\s_0 =1 \rmod \l \sp \frac{N(\s) \hspace{.03in}\s \hspace{.03in}\r(\s)}{\l}=1 \rmod \r(\s) \sp P(\s,\s_0)\hspace{.03in}N(\s_0)=1 \rmod \r(\s)
\)
\(
1\leq N(\s_0)\leq \l-1
\)
\(
N(0)=P(0,\s_0)=0; \hspace{.3in} 1\leq N(\s),P(\s,\s_0)\leq \r(\s)-1 \sp \s=1,...,\l-1
\)
\reseteqn
where $P(\s,\s_0)$ is a new integer which characterizes the transition from $\s_0$ to $\s$.  The result for $\lr(\s)$ is 
\namegroup{sectormap}
\alpheqn
\(
\lr_r^{a(j)b(l)}(\s;\s_0)=e^{\frac{\tp N(\s_0)N(\s) P(\s,\s_0) (j-l)r }{\l}}\sum_{m=0}^{\frac{\l}{\r(\s)}-1}e^{\frac{\tp N(\s_0)(j-l)\r(\s)m}{\l}}\lr_{\r(\s)m+N(\s) P(\s,\s_0) r}^{ab} \label{peridtwo}
\)
\(
r=0,...,\r(\s)-1\sp j,l=0,...,\lors-1\sp\sigma=0,...,\l-1
\)
\reseteqn
where we have explicitly indicated the $\s_0$ dependence of $\lr(\s)$ inherent in this procedure\footnote{It can be seen from (\ref{sectormap}) that different choices of $\s_0$ produce a reordering of the twisted sectors and different forms for $L$ in the untwisted sector.  These $L$'s differ however only by transformations in $S_\l\subset Aut(\gh)$, and so are physically equivalent.  This phenomenon underlies the otherwise mysterious fact that the number $n_C$ in (\ref{nCOVME}) is the same for the cyclic OVME (\ref{COVMEc}) and the reduced VME (\ref{redVME}) of the $\z_\l$(permutation)-invariant CFT's.}.  

When $\r(\s)=\l$ (so that $j=l=0$ and $P(\s,\s_0)=\s_0$), Eq. (\ref{sectormap}) relates any other solutions at order $\l$ of the cyclic OVME to the given solution at order $\l$
\(
\lr_r^{ab}(\s;\s_0)=\lr_{N(\s) \s_0 r}^{ab} \ .
\)
These relations show that all solutions of the cyclic OVME at order $\l$ are related by transformations in $Aut(\z_\l)\subset Aut(\gh)$.  This is the $Aut(\z_\l)$ covariance discussed in Ref.~\rf{5}.  Similarly, all sectors with common order $\r(\s)$ are related by transformations in $Aut(\z_{\r(\s)})$.

\subsection{Example: The $\mathbf{G_{\textup{\diag}(\s)}}$-invariant cyclic orbifolds} \label{Gdiagsec}
As an example of the technology above, consider the \textit{Lie $\gb$-invariant constructions}$^{\rf{5}}$ at order $\l$ of the cyclic OVME (\ref{COVMEc}) on simple $\gb$
\namegroup{OVMELieg}
\alpheqn
\(
\hat{T}_{\s_0}(z)=\sum_{r=0}^{\l-1} \lr_r^{ab}(\s_0) :\hat{J}_{a}^{(r)}(z) \hat{J}_b^{(-r)}(z):
\)
\(
\psi^2 \lr_r^{ab}(\s_0)=\eta^{ab}l_r  \sp l_r=l_{-r} \sp a,b=1,...,\dg
\)
\(
\hat{c}=\l x \dg \sum_{r=0}^{\l-1}l_r \sp \hat{\Delta}_0=\frac{x}{4 \l} \dg \sum_{r=0}^{\l-1}l_r r(\l-r) \label{OVMELieCD}
\)
\reseteqn
(that is, $S=1$) where $\eta^{ab}$ is the inverse Killing metric of $\gb$.  All such constructions are also in the OVME.  The explicit form of the Lie $\gb$-invariant constructions is given through $\l=6$ in Ref.~\rf{5}.  Using (\ref{peridtwo}) we find the other sectors of these orbifolds
\namegroup{LieSecs}
\alpheqn
\(
\hat{T}_\s (z)=\sum_{r,j,l} \lr_r^{a(j) b(l)}(\s) :\hat{J}_{a j}^{(r)}(z) \hat{J}_{bl}^{(-r)}(z):
\)
\(
\psi^2 \lr_r^{a( j)b( l)}(\s)  \equiv \eta^{ab}  l_r^{jl} (\s)
\)
\(
l_r^{jl}(\s)  =  e^{\frac{\tp N_0 N P r(j-l)r}{\l}}\sum_{m=0}^{\frac{\l}{\r(\s)}-1}e^{\frac{\tp N_0(j-l)\r(\s)m}{\l}} l_{\r(\s)m+N P r} 
\)
\(
\lr_r^{a(j)b(l)}(\s)  =  \lr_{-r}^{b(l)a(j)}(\s)
\)
\(
N_0\equiv N(\s_0) \sp N\equiv N(\s) \sp P\equiv P(\s,\s_0)
\)
\(
r=0,...,\r(\s)-1\sp j,l=0,...,\lors-1 \sp \gb_{\r(\s)}=\oplus_{j=0}^{\lors-1}\gb_{\r(\s)}^j
\)
\reseteqn
where we have suppressed the explicit $\s_0$ dependence of $\lr(\s)$ for simplicity.  These solutions do not generically have the $a(\cdot)\lra b(\cdot)$ symmetry in (\ref{ovmesym}):
\(
\lr_r^{a(j)b(l)}\neq\lr_r^{b(l)a(j)}
\)
so these sectors are not generically in the OVME.  

The sectors of these orbifolds have a residual Lie symmetry group $G_{\textup{\diag}(\s)}$
\namegroup{gdiaggroup}
\alpheqn
\(
 \lr_r^{c(j)d(l)}(\s) \omega_{c}^{\ a} \ \omega_{d}^{\ b} = \lr_r^{a(j)b(l)}(\s)
\sp \forall\ \omega_{a}^{\ b} \in G_{\textup{\diag}(\s)}
\)
\(
Q_a^{\diag(\s)}\equiv\sum_{j=0}^{\lors-1}\hat{J}_{aj}^{(0)}(m=0) \label{gdiagJ}
\)
\reseteqn
with Lie algebra $\gb_{\diag (\s)}$, generated by $Q_a^{\diag(\s)}$ in (\ref{gdiagJ}).

For these orbifolds we have $c=\hat{c}(\sigma)=\hat{c}$ in (\ref{OVMELieCD}) and one finds 
\alpheqn
\(
\hat{\Delta}_0(\s)=\ \l x \dg \sum_r \sum_j l_r^{jj} \frac{r(\r(\s)-r)}{4 \r^2(\s)}\hsp{3.2}
\)
\(
\hsp{.666} =
\left\{
\begin{array}{cl}
0&,\ \s=0\\
\frac{\l^2}{4\r^3(\s)}x\dg{\displaystyle{\sum_{r=0}^{\r(\s)-1} \sum_{j=0}^{\frac{\l}{\r(\s)}-1}}} l_{\r(\s) j + N(\s) P(\s)r}\hsp{.05}r(\r(\s)-r)
&,\ \s=1,...,\l-1
\end{array}
\right.
\) 
\reseteqn
for the conformal weights of the ground states $|0\rangle_\sigma$.

To identify these orbifolds, consider the untwisted sectors (with $\r(0)=1$ and $N=N(0)=0$) for which  Eq. (\ref{LieSecs}) gives
\namegroup{gdiaggroupuntwisted}
\alpheqn
\(
T(z) \equiv \hat{T}_{\s=0}(z)=\sum_{J,L} L_{J-L}^{ab}:J_{aJ}(z) J_{bL}(z):
\)
\(
L_{J-L}^{ab}\equiv\lr_0^{a(J)b(L)}(\sigma=0)=\eta^{ab} \sum_{r=0}^{\l-1}e^{\frac{2 \pi i N_0 (J-L)r}{\l}} l_r \label{VMELiegL}
\)
\(
L_{J-L}^{ab}=L_{J-L}^{ba}=L_{L-J}^{ab}=L_{J-L\pm\l}^{ab} \label{dzLie}
\)
\(
J,L=0,...,\l-1 \sp \gh=\oplus_{I=0}^{\l-1} \gb^I.
\)
\reseteqn
It follows from the symmetries in (\ref{dzLie}) that $L$ solves the reduced VME (\ref{DVME} \ref{DVMEeq},\ref{DVMEsymm}) of the $\dl$(permutation)-invariant CFT's.  This fact and the symmetry $G_{\textup{\diag}(\s)}$ in (\ref{gdiaggroup}) suggest the name $G_{\diag (\s)}$-\textit{invariant cyclic orbifolds} for this class of CFT's: These are the orbifolds in $A(\dl)/\z_\l$ whose untwisted sectors $L$ are also invariant under the symmetry $G_{\textup{\diag}}$ generated by
\vspace{-.2in}
\(
Q_a^\diag \equiv  Q_a^{\diag(0)}=\sum_{J=0}^{\l-1} J_{aJ}(m=0). 
\)
We emphasize however that the $G_{\diag (\s)}$-invariant cyclic orbifolds have the $G_{\diag(\s)}$-invariance (\ref{gdiaggroup}) in every sector.

These examples also illustrate an important aspect of our new duality transformations: Solutions such as (\ref{OVMELieg}) whose form$^{\rf{5}}$ is relatively simple\footnote{With (\ref{gdiaggroupuntwisted}) one may construct the untwisted sectors of the orbifolds corresponding to the Lie $\gb$-invariant constructions at $\l=5$ and 6 in Ref.~\rf{5}.  At $\l=5$, these are quasi-rational orbifolds with irrational conformal weights and rational central charge.  Other quasi-rational CFT's are discussed in Ref.~\rf{3}.} at the OVME level can result in intricate solutions such as (\ref{VMELiegL}) at the VME level and vice versa.  (E.g. the case$^{\rf{27}}$ of (simply-laced $\gb$)$^q$, which is relatively simple at the VME level, will give Fourier series in the twisted sectors.)

\newsection{General Orbifolds}
\label{gensec}
\subsection{Discussion}
In Section~\ref{mainsec}, we generally followed the dotted line in the commuting diagram shown in Fig.~3,

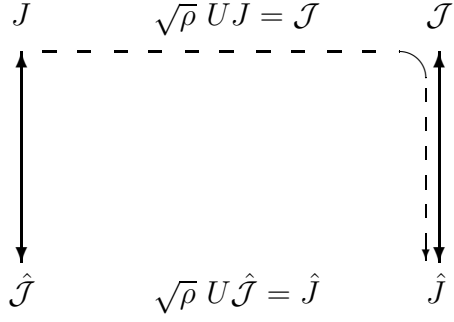
\begin{picture}(330,250)(0,0)
\put(151,220){$J$}
\put(163,210){\line(1,0){5}}
\put(190,210){\line(1,0){5}}
\put(217,210){\line(1,0){5}}
\put(244,210){\line(1,0){5}}
\put(271,210){\line(1,0){5}}
\put(176,210){\line(1,0){5}}
\put(203,210){\line(1,0){5}}
\put(230,210){\line(1,0){5}}
\put(257,210){\line(1,0){5}}
\put(284,210){\line(1,0){5}}
\put(298,200){\oval(20,20)[tr]}
\put(308,196){\line(0,-1){5}}
\put(308,185){\line(0,-1){5}}
\put(308,174){\line(0,-1){5}}
\put(308,163){\line(0,-1){5}}
\put(308,152){\line(0,-1){5}}
\put(308,141){\vector(0,-1){10}}
\put(205,220){$\sr \ UJ=\j$}
\thicklines
\put(155,210){\vector(0,-1){80}}
\put(155,130){\vector(0,1){80}}
\put(150,115){$\hat{\j}$}
\put(308,220){$\j$}
\put(205,115){$\sr \ U\hat{\j}=\hat{J}$}
\put(313,210){\vector(0,-1){80}}
\put(313,130){\vector(0,1){80}}
\put(308,115){$\hat{J}$}
\put(62,79) {$J$}
\put(74,79) {= affine Lie algebra: trivial monodromy, mixed under automorphisms}
\put(60,65) {$\j$}
\put(74,65) {= eigencurrents: trivial monodromy, diagonal under automorphisms}
\put(62,51) {$\jh$}
\put(74,51) {= twisted currents with diagonal monodromy}
\put(60,37) {$\hat{\j}$}
\put(74,37) {= twisted currents with mixed monodromy}
\put(130,11) {Fig. 3: The currents of a general orbifold}
\end{picture}

\noindent
where double arrows ($\lra$) indicate OPE isomorphisms.  In this section we follow the same path (up to normalization) for each twisted sector of the general orbifold
\(
\frac{A(H)}{H}\sp H\subset Aut(g)
\)
where $A(H)$ is any $H$-invariant CFT on $g$ (see Ref.~\rf{4}).  Here, $H$ can be any particular finite subgroup of inner or outer automorphisms of the ambient Lie algebra $g$, which itself may be simple or semisimple.  Formally, $H$ can be extended to Lie groups.
\subsection{General twisted current algebra and $A(H)/H$} \label{gencurrentsec}
The automorphisms $H$ of the untwisted affine algebra on $g$ are defined as  
\alpheqn
\(
J_a(z) J_b (w) = \frac{G_{ab}}{(z-w)^2} + \frac{i f_{ab}^{\ \ c}J_c(w)}{z-w} + O((z-w)^0) \label{origalg}
\)
\(
J_a (z)\p = \w (h_\s)_a^{\ b} J_b(z), \hspace{.3in} h_\s \in H\subset Aut(g)
\)
\(
a,b=1,...,\textup{dim}g
\)
\reseteqn
where $\{\w(h_\s)\}$ is a representation of $H$ and $J\p$ satisfies the same algebra as $J$.  Then the $H$-invariant CFT's $A(H)$ on $g$ are described by 
\alpheqn
\(
T(z)=L^{ab}:J_a(z)J_b(z): \sp L^{ab}=L^{ba}
\)
\(
L^{cd} \w (h_\s)_c^{\ a} \w (h_\s)_d^{\ b}=L^{ab},\ \ \forall\  h_\s \in H \label{HT}
\)
\reseteqn
and this set of inverse inertia tensors $L$ is guaranteed$^{\rf{4}}$ to satisfy a consistent reduced VME (such as those in Eqs. (\ref{DVME}) or (\ref{redVME})).

In general orbifold theory$^{\rf{15},\rf{38},\rf{41}}$, the sectors of the orbifolds $A(H)/H$ correspond to the conjugacy classes of $H$.  Picking one element $h_\s \in H$ from each class, we need the solutions to the $H$-eigenvalue problem
\alpheqn
\(
\w(h_\s)_a^{ \ b}  U^\dagger(\s)_b^{\ r \mu} = E_r(\s) U^\dagger(\s)_a^{\ r \mu} \sp E_r(\s)=e^{-\frac{\tp n(r)}{\r(\s)}} \label{gevprob}
\)
\(
U^\dagger(\s)_a^{\ r \nu} U(\s)_{r \nu}^{\ \ b} = \delta_a^{\ b}\sp
U(\s)_{r \mu}^{\ \ a} U^\dagger(\s)_a^{\ s \nu} = \delta_r^{\ s} \delta_{\mu}^{\ \nu} 
\)
\reseteqn
for each sector $\s$, where $\omega$ and $U$ are unitary, \{$n(r)$\} is a set of $H$-dependent integers and $\r(\s)$ is the order of $h_\s$.  The case $H=\z_\l$(permutation) was solved in Subsec. \ref{epsec}.  Following the conventions above, $r$ is the spectral index and $\mu$ (which in general depends on $r$) replaces ($a,j$) as the degeneracy index of the eigenvalue problem.

The eigencurrents $\j$ of sector $\s$ are then defined as 
\alpheqn
\(
\j_{r \mu}(z) \equiv \chi(\s)_{r \mu} U(\s)_{r \mu}^{\ \ a} J_a(z), \hspace{.3in} J_a(z)= U^\dagger(\s)_a^{\ r \mu} (\chi(\s)_{r \mu})^{-1}\j_{r \mu}(z) \label{geneigenj}
\)
\(
\j_{r\mu}(z)\p=E_r(\s) \hsp{.03}\j_{r \mu}(z) \sp \j_{r \mu}(z e^{\tp})=\j_{r \mu}(z)
\)
\reseteqn
where $\chi(\s)_{r \mu}$ is an arbitrary normalization\footnote{In the previous sections (and in Fig.~3) we chose $\chi(\s)_{r \mu}=\sqrt{\r(\s)}$ to obtain the conventional form (\ref{twistalg}) of orbifold affine algebra for $\hat{J}$ in each sector.}.  

The OPE's of the eigencurrents $\j$ are easily computed and then one uses the OPE isomorphism 
\(
\j_{r \mu}(z) \sa \hat{J}_{r \mu}(z)
\)
to find the algebra $\gb\equiv\gb(H\subset Aut(g);\s)$ of the twisted currents $\hat{J}$ with diagonal monodromy for each sector $\s$:
\namegroup{gendiagJ}
\alpheqn
\(
\hat{J}_{r \mu}(z) \hat{J}_{s \nu} (w) = \frac{\g_{r \mu, s \nu}(\s)}{(z-w)^2} + \frac{i \fr_{r \mu, s \nu}^{\ \ \ \ \ \ t \delta}(\s) \hat{J}_{t \delta}(w)}{z-w} + O((z-w)^0)
\)
\(
\hat{J}_{r \mu}(ze^{\tp}) = E_r(\s) \hat{J}_{r \mu}(z) \sp E_r(\s)=e^{-\frac{\tp n(r)}{\r(\s)}}
\)
\ba
\g_{r \mu, s \nu}(\s) &\equiv& \chi(\s)_{r \mu} \chi(\s)_{s \nu} U(\s)_{r \mu}^{\ \ a} U(\s)_{s \nu}^{\ \ b} G_{ab}\\
\fr_{r \mu, s \nu}^{\ \ \ \ \ t \delta}(\s) &\equiv& \chi(\s)_{r \mu} \chi(\s)_{s \nu}  U(\s)_{r \mu}^{\ \ a} U(\s)_{s \nu}^{\ \ b} f_{ab}^{\ \ c}   U^\dagger(\s)_c^{\ \ t \delta} (\chi(\s)_{t \delta})^{-1}.
\ea
\reseteqn
\textbf{The result (\ref{gendiagJ}) is presumably the most general twisted current algebra, since it collects the twisted current algebras of every sector of every orbifold} ${\mathbf{A(H)/H}}$, ${\mathbf{H\subset Aut}}${\textbf{(}}$g$\textbf{)}.  See Eqs. (\ref{twistalg}) and (\ref{grho}) for the special case $\gb_{\r(\s)} \equiv \gb(\zl \subset Aut(g);\s)$, which is the correct set of orbifold affine algebras for \taz.  

Note in particular that the OPE form of the general twisted current algebra (\ref{gendiagJ}) can be obtained from the untwisted affine algebra (\ref{origalg}) by the duality algorithm
\(
a\sat r\mu \sp G \sat \g \sp f \sat \fr \sp J \sat \hat{J}. \label{indicesJ}
\)
We will return below to the representation theory of the algebras (\ref{gendiagJ}).

To obtain the stress tensors of the twisted sectors, one uses (\ref{geneigenj}) to eliminate $J$ in favor of $\j$ in the $H$-invariant stress tensor $T$ given in (\ref{HT}).  Then, using the correspondence
\(
T(z)\ \sa\ \hat{T}_\s (z) 
\)
one finds \textbf{the general duality transformations} $L\rightarrow\lr$
\namegroup{HTgroup}
\alpheqn
\(
\hat{T}_\s (z)=\lr^{r \mu, s \nu}(\s):\hat{J}_{r \mu}(z) \hat{J}_{s \nu}(z):
\)
\(
\lr^{r \mu, s \nu}(\s)=L^{ab} U^\dagger(\s)_a^{\ \ r \mu} U^\dagger(\s)_b^{\ \ s \nu}(\chi(\s)_{r \mu})^{-1}(\chi(\s)_{s \nu})^{-1}=\lr^{s \nu, r \mu}(\s) \label{Halpern}
\)
\(
\hat{c}(\s)=c=2G_{ab}L^{ab}
\)
\reseteqn
\textbf{which give the stress tensors $\mathbf{\hat{T}_\s}$ of all the twisted sectors of all the orbifolds} $\mathbf{A(H)/H}$, $\mathbf{H\subset Aut}${\textbf{(}}$g$\textbf{)}.  The general duality transformations include the special case (\ref{mainres}) when $H=\z_\l$(permutation).  Note also that the general duality transformations in (\ref{Halpern}) are always invertible ($L\lra\lr$) so that one may solve for the inverse inertia tensor of all sectors in terms of the inverse inertia tensor of any given sector (see e.g. Eq.~(\ref{sectormap})). 

The twisted sectors (\ref{HTgroup}) show K-conjugation covariance$^{\rf{12},\rf{18}-\rf{23},\rf{1},\rf{3}}$ in the form
\namegroup{genK}
\alpheqn
\getletter{genKa}
\vspace{-.1in}
\(
\hat{T}_{\s} (z)_{\sgb}=\hat{T}_\s (z)+\tilde{\hat{T}\hspace{.035in}}_{\hspace{-.05in}\s} (z)
\)
\vspace{-.3in}
\getletter{genKb}
\(
\tilde{\hat{T}\hspace{.035in}}_{\hspace{-.05in}\s} (z) \hat{T}_\s (w)=\reg
\)
\vspace{-.3in}
\getletter{genKc}
\(
\lr_{\sgb}^{r \mu, s \nu}(\s)=\lr^{r \mu, s \nu}(\s)+\tilde{\lr}^{r \mu, s \nu}(\s) \label{genKconjtwist}
\)
\vspace{-.3in}
\getletter{genKd}
\(
L_g^{ab}= L^{ab}+\tilde{L}^{ab} \label{genKconjuntwist}
\)
\reseteqn
where the orbifold $K$-conjugation (\ref{genK} \ref{genKa}-\ref{genKc}) is the map, via the duality transformation (\ref{Halpern}), of the ordinary $K$-conjugation (\ref{genK}\ref{genKd}) in the untwisted sector.  Here $L_g$ is the inverse inertia tensor of the affine-Sugawara construction$^{\rf{12},\rf{18},\rf{19}-\rf{21}}$ on $g$ and $\hat{T}_{\s}(z)_{\sgb}$ with $L_g\rightarrow\lr_{\sgb}$ is the general orbifold affine-Sugawara construction on the twisted current algebra $\gb=\gb(H\subset Aut(g);\s)$ in (\ref{gendiagJ}).  

Note also the identity 
\(
0=\lr^{r \mu, s \nu}(\s) \ (1-E_r(\s) E_s(\s))=\lr^{r \mu, s \nu}(\s) \ (1-e^{-\frac{\tp(n(r)+n(s))}{\r(\s)}})
\)
which follows from (\ref{Halpern}), the $H$-symmetry of $L^{ab}$ in (\ref{HT}) and the $H$-eigenvalue problem (\ref{gevprob}).  This tells us that the stress tensors of the twisted sectors have trivial monodromy
\alpheqn
\(
\lr^{r \mu, s \nu}(\s) \propto \delta_{n(r)+n(s),0 \mod \r(\s)} \label{trivmon}
\)
\(
\hat{T}_\s (ze^{\tp})=\hat{T}_\s (z)
\)
\reseteqn
as required.  The relation in (\ref{trivmon}) generalizes the form (\ref{lrnomono}) found above for \taz.

We also give \textbf{the A(H)/H orbifold Virasoro master equation, which collects the inverse inertia tensors of A(H)/H, H $\subset$ Aut($g$) for each H and $\mathbf{\s}$:}
\alpheqn
\ba
& &\lr^{r \mu, s \nu}(\s) =2 \lr^{r \mu, r' \mu'}(\s) \g_{r' \mu',s' \nu'}(\s) \lr^{s' \nu', s \nu}(\s) \nonumber\hsp{.6}\\
& &\hsp{.8}  - \lr^{r' \mu', s' \nu'}(\s) \lr^{t \eta, t' \eta'}(\s) \fr_{r' \mu', t \eta}^{\ \ \ \ \ \ \ r \mu}(\s) \fr_{s' \nu', t' \eta'}^{\ \ \ \ \ \ \ \ s \nu} (\s)\nonumber \hsp{.6}\\
& &\hsp{.8} - \lr^{r' \mu', s' \nu'}(\s) \fr_{r' \mu', t \eta}^{\ \ \ \ \ \ \ t' \eta'}(\s) \fr_{s' \nu', t' \eta'}^{\ \ \ \ \ \ \ \ (r \mu} (\s) \lr^{s \nu), t \eta}(\s)
\hsp{.20}\textup{(dual to VME)}\hsp{.6} \label{HOVME}\\
&&\hsp{.4}\lr^{r \mu, s \nu}(\s)=\lr^{s \nu, r \mu}(\s) \hspace{2.29in} (\textup{dual to $L^{ab}=L^{ba}$})\hsp{.6}\\
&&\hsp{.4}\lr^{r \mu, s \nu}(\s) \ (1-E_r(\s) E_s(\s))=0 \hspace{1.0996in} (\textup{dual to $L_H\hsp{.02} \w_H \hsp{.02} \w_H=L_H$})\hsp{.6}
\ea
\reseteqn
where $\g(\s)$ and $\fr(\s)$ are the twisted metric and twisted structure constants in (\ref{gendiagJ}), and the round brackets mean $r \mu \leftrightarrow s \nu$ symmetrization as in the VME.  For \taz, $H=\zl\subset Aut(g)$, this system reduces to the cyclic OVME on $\gb_{\r(\s)}$ in (\ref{covmeg}) and the symmetries of the cyclic OVME in (\ref{COVMEc}).  

The general OVME (\ref{HOVME}) can be derived by the general duality transformations (\ref{HTgroup}) from the VME (\ref{VMEa}).  Alternately, the general OVME can be obtained immediately from the VME by the duality algorithm
\(
a\sat r\mu\sp G \sat \g \sp f \sat \fr \sp L \sat \lr.
\)
The reason this works is that the OPE form of the general twisted current algebra satisfies the same duality algorithm (\ref{indicesJ}), so that every step of a direct OPE verification of the general OVME follows immediately from the corresponding steps for the VME.  Note in particular that, under the index map $a\longrightarrow r \mu$, the indices in the general OVME are threaded in exactly the same way as those in the VME.

This completes our general discussion of the new duality transformations in orbifold theory but there are several remaining loose ends.  First, one may use the results above to explicitly verify that only one element of each conjugacy class is necessary for non-abelian orbifolds: When $\w$ and $\w\p$ are in the same conjugacy class of $H$, follow the steps
\alpheqn
\(
\w\p=v^\dagger\w v\sp v^\dagger v=1 \sp v\in H\subset Aut(g)
\)
\(
v_a^{\ c} v_b^{\ d}G_{cd}=G_{ab}\sp 
v_a^{\ d} v_b^{\ e} (v^\dagger)_g^{\ c} f_{de}^{\ \ g}=f_{ab}^{\ \ c}
\)
\(
L^{cd}(v^\dagger)_c^{\ a}(v^\dagger)_d^{\ b}=L^{ab}\sp 
E\p=E\sp U\p=Uv
\)
\(
\g(U\p)=\g(U)\sp\fr(U\p)=\fr(U)\sp\lr(U\p)=\lr(U).
\)
\reseteqn
It follows that $\g, \fr$, and $\lr$ are class functions, so that the general twisted current algebra (\ref{gendiagJ}) and the stress tensors (\ref{HTgroup}) are invariant across each conjugacy class.  As seen in the example of App.~C, we expect that the ambiguities in $U$ at fixed $\w$ correspond to automorphisms of the general twisted current algebra.

Second, one may easily construct the currents $\hat{\j}$ with mixed monodromy
\alpheqn
\(
\hat{\j}_a(z) \equiv U^\dagger(\s)_a^{\ \ r \mu} (\chi(\s)_{r \mu})^{-1} \hat{J}_{r \mu}(z) \sp \hat{\j}_a(ze^{\tp})=\omega (h_\s)_a^{\ b}\hat{\j}_b(z)
\)
\(
\hat{J}_{r \mu}(z) = \chi(\s)_{r \mu} U(\s)_{r \mu}^{\ \ a} \hat{\j}_a(z)
\)
\reseteqn
whose OPE's are isomorphic to the original current algebra (\ref{origalg}).

Third, we emphasize that our result (\ref{HTgroup}) also contains more general orbifolds of the form
\(
\frac{A(H\p)}{H} \subset \frac{A(H)}{H} \sp H \subset H\p \subset Aut(g)
\)
because $A(H\p)\subset A(H)$.  In such cases, one uses the same duality transformations (\ref{HTgroup}) for $H$, but now with the inverse inertia tensors $L_{H\p}$
\(
L_{H\p} \hspace{.03in} \w_{H\p} \hspace{.03in} \w_{H\p}=L_{H\p}\subset L_H
\)
of the $H\p$-invariant CFTs (see Subsec.~\ref{OVMEsec}).  It follows that the number of sectors of each $A(H\p)/H$ is still the number of conjugacy classes of $H$.

A more practical matter is the representation theory of the general twisted current algebra $\gb(H\subset Aut(g);\s)$ in (\ref{gendiagJ}).  When $g$ is simple, these algebras are inner- or outer-automorphically twisted affine Lie algebras, depending on whether $H$ is a group of inner or outer automorphisms of $g$.  The representation theory of these two types of twisted algebras is discussed in Refs.~\rf{43}-\rf{45}, \rf{17} and Refs.~\rf{46}, \rf{42} respectively.  

When $g$ is semisimple and $H\subset S_\l$ acts as (outer-automorphic) permutations among $\l$ copies, we expect that the twisted current algebras (\ref{gendiagJ}) are commuting sets of orbifold affine algebras at various orders $\r$, whose representation theory was obtained via the orbifold induction procedure in Ref.~\rf{9}.  This is implicit in the work of Ref.~\rf{41} and follows because the elements of all the subgroups of $S_\l$ can be written as products of disjoint cyclic permutations.  As an example, this conclusion is illustrated for the permutation orbifolds $A(S_3)/S_3$ in the following subsection.

When $g$ is semisimple and $H$ is any subgroup of $S_\l$ times an inner automorphism of $g$, then one expects that (\ref{gendiagJ}) gives various versions of the ``doubly-twisted'' orbifold affine algebras (which combine inner- and outer- automorphic twists) obtained in Ref.~\rf{5}.  The representation theory of these algebras should involve a synthesis of the principles already known for inner automorphisms and for outer automorphisms by permutation.

Finally, we note that when $A(H)$ has a larger chiral algebra, such as a superconformal algebra or a non-linear chiral algebra, the same techniques can be used to treat the larger twisted chiral algebras in each sector of $A(H)/H$.

\subsection{Example: the permutation orbifolds $A(S_3)/S_3$} \label{S3sec}

As an example of the general development above we consider the permutation orbifolds $A(S_3)/S_3$, starting with the $S_3$(permutation)-invariant CFT's $A(S_3)$ on $\gb \oplus \gb \oplus \gb$,
\namegroup{symanz}
\alpheqn
\(
T=L^{a(J)b(L)}:J_{aJ}J_{bL}:\sp
L^{a(J)b(L)}=\delta_{JL}\l^{ab}+l^{ab} 
\)
\getletter{symanzL}
\(
\arange\sp J,L=0,1,2
\)
\reseteqn
where $\l^{ab}$ and $l^{ab}$ are $a\hsp{-.06}\lra\hsp{-.06} b$ symmetric.  The $S_3$(permutation)-invariant CFT's satisfy a consistent reduced VME which we omit for brevity.

First, pick one element $\w$ from each conjugacy class of $S_3$,
\(
\w_0=\left(
\begin{array}{ccc}
1&0&0\\0&1&0\\0&0&1
\end{array}
\right)
\hsp{.5}
\w_1=\left(
\begin{array}{ccc}
0&0&1\\1&0&0\\0&1&0
\end{array}
\right)
\hsp{.5}
\w_2=\left(
\begin{array}{ccc}
1&0&0\\0&0&1\\0&1&0
\end{array}
\right)
\)
where $\w_0$ corresponds to the untwisted sector with $T_{\w_0}=T$ in (\ref{symanz}).  The element $\w_1$ is also an element of $\z_3$, so the sector corresponding to $\w_1$ is included in the discussion of $\z_3$ orbifolds above  (but with the $S_3$-invariant form in (\ref{symanz}\ref{symanzL}}) instead of the $\z_3$-invariant form (\ref{Lcyc})).  The element $\w_2$ is new, but it is a direct sum of an element of $\z_1$ and an element of $\z_2$.

Using the duality transformation (\ref{themap}) we find the stress tensor $\hat{T}_{\w_1}$ of the twisted sector corresponding to $\w_1$,
\alpheqn
\(
\hat{T}_{\w_1}=(l^{ab}+\frac{\l^{ab}}{3}):\jh^{(0)}_a\jh^{(0)}_b:+\frac{2\l^{ab}}{3}:\jh^{(1)}_a\jh^{(-1)}_b:
\)
\(
\hat{J}_a^{(r)}((m+\frac{r}{3})\geq0)|0\rangle=0
\)
\(
\hat{c}(\s)=c=6k\eta_{ab}(\l+l)^{ab} \sp \hat{\Delta}_0(\s)=\frac{2k}{9}\eta_{ab}\l^{ab}
\)
\reseteqn
where $\jh^{(r)}_a$, $r=0,1,2$ is an orbifold affine algebra of order 3 on simple $\gb$ at level $\hat{k}=3k$.

For the new sector corresponding to $\w_2$, the general discussion of Subsec.~\ref{gencurrentsec} gives
\alpheqn
\(
\hat{T}_{\w_2}=\l^{ab}:J_aJ_b+\frac{1}{2}(\hat{J}_a^{(0)}\hat{J}_b^{(0)}+\hat{J}_a^{(1)}\hat{J}_b^{(-1)}): \ + \ l^{ab}:J_aJ_b+J_a\hat{J}_b^{(0)}+\hat{J}_a^{(0)}J_b+\hat{J}_a^{(0)}\hat{J}_b^{(0)}:
\)
\(
\hat{J}_a^{(r)}((m+\frac{r}{2})\geq0)|0\rangle=J_a(m\geq0)|0\rangle=0
\)
\(
\hat{c}(\s)=c=6k\eta_{ab}(\l+l)^{ab} \sp \hat{\Delta}_0(\s)=\hat{\Delta}_0=\frac{k}{8}\eta_{ab}\l^{ab}.
\)
Here $J_a$ is an affine algebra (i.e. an orbifold affine algebra at order 1) on simple $\gb$ at level $k$, and $\hat{J}_a^{(r)}$ with $r=0,1$ is an orbifold affine algebra of order 2 on simple $\gb$ at level $\hat{k}=2k$.  The appearance of orders 1 and 2 in this sector correlates directly with the $\z_1$ and $\z_2$ decomposition of $\w_2$.  The eigencurrents of this sector show that the currents $J_a$ and $\hat{J}_a^{(r)}$ commute, in accord with the general remarks above.  
\bigskip

\noindent
{\bf Acknowledgements} 

For helpful discussions, we thank L. Borisov, J. Fuchs, E. Kiritsis, N. Obers, C. Schweigert, V. Serganova and E. Verlinde.  

J. E. and M. B. H. were supported in part by the Director, Office of Science, Office of High Energy and Nuclear Physics, Division of High Energy Physics, of the U.S. Department of Energy under Contract DE-AC03-76SF00098 and in part by the National Science Foundation under grant PHY95-14797.

\appendices
\subsection{Useful Identities}
\(
\sum_{r=0}^{\r-1}e^{-\frac{\tp rs}{\r}}=\sum_{r=0}^{\r-1}e^{-\frac{\tp N rs}{\r}}=\r \ \delta_{s,0 \mod \r} \label{perid}
\)
\(
\delta_{j,J+\sigma \mod \frac{\l}{\r}} = \delta_{j,J \mod \frac{\l}{\r}} 
\label{Jper}
\)
\(
\delta_{j,J \mod \frac{\l}{\r}} \ \ \delta_{l, J \mod \frac{\l}{\r}}= \delta_{l,j \mod \frac{\l}{\r}} \ \ \delta_{l, J \mod \frac{\l}{\r}} \label{sumrule}
\)
\(
\sum_{t=0}^{\r-1} \delta_{t, m-n \mod \r}=1 \label{faddeev}
\)
\alpheqn
\agetletter{halftricklett}
\begin{displaymath}
\hspace{-2in} \sum_{r,s=0}^{\r-1} \sum_{J,L=0}^{\l-1} f(J-L) \ e^{\frac{\tp N r(j-J)}{\l}} \ e^{\frac{\tp N s(l-L)}{\l}} \delta_{j, J \mod \frac{\l}{\r}} \ \ \delta_{l, L \mod \frac{\l}{\r}} \nonumber\\
\end{displaymath}
\ba
& = & \sum_{r,s=0}^{\r-1} \sum_{n,m=0}^{\r-1} f(j+\frac{\l}{\r}m - l - \frac{\l}{\r}n) \ e^{-\frac{\tp N rm}{\r}} \ e^{-\frac{\tp N sn}{\r}} \nonumber \\
& = & \sum_{r,s=0}^{\r-1} \sum_{n,m=0}^{\r-1} \sum_{t=0}^{\r-1} \delta_{t,m-n \mod \r} \ \ f(j-l +\frac{\l}{\r}(m - n)) \ e^{-\frac{\tp N (rm+sn)}{\r}} \nonumber \\
& = & \r \sum_{r,t}^{\r-1} \sum_{s=0}^{\r-1} f(j-l+\frac{\l}{\r}t) e^{-\frac{\tp N rt}{\r}} \delta_{r+s, 0 \mod \r} \label{halftrick} \\
& = & \r \sum_{r,t=0}^{\r-1} f(j-l+\frac{\l}{\r}t) \ e^{-\frac{\tp N rt}{\r}}  \label{trick}
\ea
\reseteqn
\(
L_0^{ab}=\sum_r \lr_r^{a(j)b(j)}(\s)=\frac{\r(\sigma)}{\l}\sum_{j,l,r}\delta_{jl}\lr_r^{a(j)b(l)}(\s) \label{lzero}
\)
\(
\lr_r^{a(j)b(l)}(\s)=\lr_{r,-r}^{a(j)b(l)}(\s)=\frac{1}{\r^2(\s)}\sum_{J,L=0}^{\l-1}L_{J-L}^{ab}e^{\frac{\tp N(\s) r[(j-l)-(J-L)]}{\l}}\delta_{j,J \mod \lors} \delta_{l,L \mod \lors} \label{loldL}
\)
Here, $N=N(\s)$ is relatively prime to $\r=\r(\s)$ (see App.~B).  The identity (\ref{faddeev}) is a ``Faddeev-Popov'' insertion used to obtain (\ref{halftrick}).
\subsection{The integer $N(\s)$} \label{appN}
As examples of the definitions in Subsec~\ref{Tofsig}:

\hsp{.3}
$\l=3$:
\hspace{.1666in}{\begin{tabular}{r|l|l|l}
$\s$&$\r$&$M$&$N$\\ \hline
0&1&\ 0&0\\
1&3&\ 1&1\\
2&3&\ 2&2\\
\end{tabular}
\hsp{1.666}
$\l=4$:
\hspace{.1666in}{\begin{tabular}{r|l|l|l}
$\s$&$\r$&$M$&$N$\\ \hline
0&1&\ 0& 0\\
1&4&\ 1& 1\\
2&2&\ 1& 1\\
3&4&\ 3& 3\ \ \ .\\
\end{tabular}}
\vspace{.1666in}
\\
\noindent
For prime $\l$ all twisted sectors have $\r(\s)=\l$ and therefore $M(\s)=\s$.  Also $N(0)=0$, $N(1)=1$ and $N(\l-\s)=\r(\s)-N(\s)$ for $\s \neq 0$. Among the elements with $\r=\l$, only $\s=1$ has $N=1$.

From the definition of $N(\s)$ in (\ref{defints}) and a well-known theorem by Euler, one finds that 
\alpheqn
\(
N(\s)=(\frac{\s \r(\s)}{\l})^{\phi(\r(\s))-1} \rmod \r(\s) \sp 1\leq N(\s) \leq \r(\s)-1
\)
\(
\s=1,...,\l-1
\)
\reseteqn
where $\phi(\r)$ is the number of elements of order $\r$ in $\z_\l$ (or equivalently in $\z_\r$).

\subsection{Automorphisms of the Orbifold Affine Algebra}

The general solution $U$ of the eigenvalue problem (\ref{eprob}) is easily found,
\namegroup{genU}
\alpheqn
\(
U_{rj;J}=\sum_{l=0}^{\lor-1}V_{rl;J} \ A_{lj}(r) =\isr e^{\frac{\tp r\b(j)}{\r}} \sum_{l=0}^{\lor-1}P_{lj}e^{-\frac{\tp Nr(l-J)}{\l}}\delta_{l,J\mod\lor}
\)
\(
\ud_{J;rj}=\sum_{l=0}^{\lor-1} V_{-r,l;J} \ A_{lj}(-r) =\isr e^{-\frac{\tp r\b(j)}{\r}} \sum_{l=0}^{\lor-1}P_{lj} e^{\frac{\tp Nr(l-J)}{\l}}\delta_{l,J\mod\lor} 
\)
\reseteqn
for each sector $\s$, where $E_r$ in (\ref{evconv}) is unchanged, $U$ is still unitary and
\alpheqn
\(
A_{jl}(r)=P_{jl}e^{\frac{\tp  r\b(l)}{\r}}=A^*_{jl}(-r)=(A^{-1})_{lj}(-r)
\)
\(
V_{rl;J}=\isr e^{-\frac{2 \pi iNr(l-J)}{\l}}\delta_{l,J\mod\lor}
\)
\(
\b(j)\in [0,\r-1] \bigcap \z.
\)
\reseteqn
The $P$'s are permutation matrices, satisfying
\namegroup{Perms}
\alpheqn
\(
P_{jl}^*=P_{jl}, \hspace{.3in} P_{jj\p}P_{lj\p}=\delta_{jl}P_{jj\p}, \hspace{.3in} \sum_{j'} P_{jj'}=\sum_j P_{jj'}=1
\)
\(
\sum_{j\p}P_{jj\p}P_{lj\p}=\sum_{j\p}P_{j\p j}P_{j\p l}=\delta_{jl}
\)
\reseteqn
but otherwise the quantities $P_{jl},\b(j)$ and hence $A_{jl}(r)$ can be chosen differently for each sector $\s$.

The $V$'s here are the $U$'s of Sec.~\ref{mainsec}, and the special case $U=V$ corresponds to $A_{jl}=\delta_{jl}$ (i.e. $P_{jl}=\delta_{jl}$ and $\beta(j)=0$).  The general $U$'s in (\ref{genU}) also satisfy (\ref{Ustar}) and (\ref{John}) and so can be used in all formulae involving $U$ and $U^\dag$ in the text. 

Following the steps of the text, we now obtain
\namegroup{mainresb}
\alpheqn
\(
\hat{T}(z)=\sum_{r,j,l}\lr_r^{a(j)b(l)}:\jh_{aj}^{(r)}(z)\jh_{bl}^{(-r)}(z):
\)
\ba
\lr_{r}^{a(j)b(l)} &=& \frac{1}{\r}\sum_s\sum_{j\p,l\p}L^{ab}_{\lor s+j\p-l\p}e^{-\frac{\tp Nrs}{\r}}A_{j\p j}(-r)A_{l\p l}(r) \\
&=& \frac{1}{\r}e^{-\frac{\tp r(\b(j)-\b(l))}{\r}} \sum_s\sum_{j\p,l\p}L^{ab}_{\lor s+j\p-l\p}e^{-\frac{\tp Nrs}{\r}}P_{j\p j}P_{l\p l}
\ea
\reseteqn
for the stress tensors $\hat{T}$ of the twisted sectors.

In fact, $A$ is an automorphism of the orbifold affine algebra (\ref{TCOPE})
\alpheqn
\(
\jh_{aj}^{(r)}(z)\p \equiv\sum_lA_{jl}(-r)\jh_{al}^{(r)}(z)
\)
\(
\jh_{aj}^{(r)}(z)\p\jh_{bl}^{(s)}(w)\p \hspace{-.03in} = \hspace{-.03in} \delta_{j l}[\frac{\r k \hspace{.016in} \eta_{ab}\delta_{r+s,0\mod\r(\sigma)}}{(z-w)^2}+\frac{if_{ab}^{\ \ c}\jh_{cj}^{(r+s)}(w)\p}{(z-w)}]+\reg 
\)
\reseteqn
and its mode form in each sector, so that the physically-equivalent set of stress tensors
\alpheqn
\(
\hat{T}(z)=\sum_{r,j,l}\lr_r^{a(j)b(l)\prime}:\jh_{aj}^{(r)}(z)\p\jh_{bl}^{(-r)}(z)\p:
\)
\(
\lr_{r}^{a(j)b(l)\prime}
=\sum_{j\p,l\p} A_{j j\p}(r) A_{l l\p}(-r) \lr_r^{a(j\p)b(l\p)}
=\frac{1}{\r}\sum_s L^{ab}_{\lor s+j-l}e^{-\frac{\tp Nrs}{\r}} \label{lprime}
\)
\reseteqn
is obtained in terms of the automorphically-equivalent currents $\hat{J}\p$. 

\subsection{Unitarity}
In this appendix, we use the results of Sec.~\ref{mainsec} to show that unitarity of the untwisted sector of any orbifold \taz \ implies unitarity of all the twisted sectors.  

Unitarity has two components in any sector.  The first component is the requirement of a good Hilbert space:  On compact $g$, the duality transformations of Sec.~\ref{mainsec} show that the invariant levels of the affine algebra (\ref{ecaa}) and the orbifold affine algebra (\ref{twistalg}) are related as follows
\(
x=\frac{2k}{\psi^2}\in\z^+\ \sa\ \hat{x}=\r x\in\r\hsp{.02}\z^+
\)
so a good Hilbert space in the untwisted sector implies a good Hilbert space in each twisted sector, as discussed in Ref.~\rf{9} and noted in Eq.~(\ref{xl}).

The second component of unitarity is $L_\s(m)^\dagger=L_\s(-m)$, where $L_\s(m)$ are the Virasoro generators of sector $\s$.  As usual$^{\rf{3}}$, we discuss this part of the problem only in Cartesian bases of the relevant Lie algebras (equivalent statements in other bases are easily deduced).

In such bases, we recall the current mode relations
\alpheqn
\(
J_{aJ}(m)^\dagger=J_{aJ}(-m)
\)
\(
\hat{J}_{aj}^{(r)} (m+\frac{r}{\r})^\dagger = \hat{J}_{aj}^{(\r-r)} (-(m+1)+\frac{\r-r}{\r})=\hat{J}_{aj}^{(-r)} (-m-\frac{r}{\r}) \label{Jmodedag}
\)
\(
\hat{J}_{aj}^{(r \pm \r)} (m+\frac{r \pm \r}{\r})=\hat{J}_{aj}^{(r)} (m \pm 1 +\frac{r}{\r}) \label{Jmodeperiod}
\)
\reseteqn
given in Ref.~\rf{9}, where the adjoint operation in (\ref{Jmodedag}) was derived from the orbifold induction procedure.  Eq.~(\ref{Jmodeperiod}) is the mode form of Eq.~(\ref{Jtwistperiod}).  It then follows from the mode form of $T$ and $\hat{T}$ in (\ref{redT}) and (\ref{twstress}) that ($*$ is complex conjugation)
\alpheqn
\(
(L^{a(J)b(L)})^*=L^{a(J)b(L)} \label{Lunitarity}
\)
\(
(\lr_r^{a(j)b(l)})^*=\lr_{-r}^{a(j)b(l)} \label{lrunitarity}
\)
\reseteqn
are the extra conditions on the inverse inertia tensors such that $L(m)^\dagger=L(-m)$ in all sectors.  The condition (\ref{Lunitarity}) in the untwisted sector is only a special case of the known unitarity condition $(L^{ab})^*=L^{ab}$ for any solution of the VME in a Cartesian basis.

It is easily checked from the duality transformations in (\ref{mainres} \ref{themaplett},\ref{allsymslett}) that (\ref{Lunitarity}) in the untwisted sector guarantees (\ref{lrunitarity}) in the twisted sectors.  

Moreover, unitarity of any given sector of the orbifold implies the unitarity of every sector.

\subsection{Orbifolds entirely in the OVME} \label{AppEx}
In this appendix, we study examples of the permutation orbifolds 
\(
\frac{A(\dl)}{\z_\l}
\) 
all of whose sectors are in the OVME (\ref{OVMEgrp}).  More precisely, this means that we assume the reflection symmetry in (\ref{appRsym}) of the $\dl$(permutation)-invariant CFT's
\alpheqn
\(
T=\sum_{J,L} L^{a(J)b(L)} :J_{aJ}J_{bL}:=\sum_{J,L} L_{J-L}^{ab} :J_{aJ}J_{bL}: \sp
L_{J-L}^{ab}=L_{L-J}^{ab}
\)
\(
\gh=\oplus_{I=0}^{\l-1}\gb^I \sp \gb^I \cong \gb
\)
\(
\arange \sp J,L=0,...,\l-1 \label{appRsym}
\)
\reseteqn
where $\gb$ is simple, and every current $J_{aI}$ is taken at the same level $k$.  Using the duality transformations in (\ref{themap}), we will also require the gauge/periodicity condition of the OVME in (\ref{appGsym}) 
\alpheqn
\(
\hat{T}_\s =\sum_{r,j,l} \lr_r^{a(j)b(l)}(\s) :\hat{J}_{aj}^{(r)} \hat{J}_{bl}^{(-r)}: \sp \lr_r^{a(j)b(l)}(\s)=\lr_{-r}^{a(j)b(l)}(\s) \label{appGsym}
\)
\(
\s=1,...,\l-1 \sp r=0,...,\r(\s)-1 \sp j,l=0,...,\lors-1  
\)
\reseteqn
for all the twisted sectors $\lr$.  The orbifold currents $\hat{J}$ satisfy the orbifold affine algebra $\gb_{\r(\s)}$ in (\ref{twistalg}), and the other symmetries of $L$ and $\lr$ in (\ref{DVME}) and (\ref{OVMEgrp}) are also assumed.

In all these examples, $\hat{T}_\s$ and $\hat{\Delta}_0(\s)$ can be read to include $T=\hat{T}_{\s=0}$ and $\Delta_0=\hat{\Delta}_0 (\s=0)=0$ via the identification 
\(
L_{J-L}^{ab}=\lr_0^{a(J)b(L)}(\s=0)
\)
for the untwisted sector.

\[\]
\bu{ \textbf{Cyclic copy orbifolds}}\\
An important class of cyclic orbifolds was identified in Ref.~\rf{9}, which we call the \textit{cyclic copy orbifolds}
\(
\frac{\times_{I=0}^{\l-1} A_I}{\z_\l} \subset \frac{A(S_\l)}{\z_\l} \sp \z_\l (\textup{permutation}) \subset S_\l (\textup{permutation}) \subset Aut(\gh)
\label{noninter} 
\)
because the stress tensor of the untwisted sector of each of these orbifolds is a sum of $\l$ commuting copies:
\alpheqn
\(
L_{J-L}^{ab}=L_0^{ab}(k)\delta_{JL} \label{localL}
\)
\(
T_I=L_0^{ab}(k):J_{aI}J_{bI}:\sp c_I=2k\eta_{ab}L_0^{ab}(k) 
\)
\(
T=\sum_{I=0}^{\l-1}T_I=\sum_{I=0}^{\l-1} L_0^{ab}(k):J_{aI}J_{bI}:\sp
c=\sum_{I=0}^{\l-1}c_I=2\l k\eta_{ab}L_0^{ab}(k)
\)
\(
T_I(z)T_J(w)=\reg\textup{\ when\ }I\neq J.
\)
\reseteqn
Here, each  $A_I$ is an  equivalent affine-Virasoro construction on $J_{aI}$, $\eta_{ab}$ is the Killing metric of $\gb$ and $L_0^{ab}(k)$ solves the VME on simple $\gb$ at level $k$.

The duality transformation (\ref{mainres}) gives the twisted sectors of these orbifolds 
\alpheqn
\(
\lr_r^{a(j)b(l)}(\s) = \frac{1}{\r(\s)} \delta_{jl} L_0^{ab}(k) \label{noninterl}
\)
\(
\hat{T}_\s =\frac{1}{\r(\s)}L_0^{ab}(k)\sum_{r=0}^{\r(\s)-1}\sum_{j=0}^{\lors-1}:\jh_{aj}^{(r)}\jh_{bj}^{(-r)}:
\)
\(
\hat{c}(\s)=c=2\l k \eta_{ab}L_0^{ab}(k) \sp 
\hat{\Delta}_0(\s)=\frac{\l}{12}k\eta_{ab}L_0^{ab}(k)(1-\frac{1}{\r^2(\s)})
\)
\reseteqn
where $\r(\s)$ is the order of $h_\s$.  

Note that the inverse inertia tensor $L$ is maximally localized in $K=J-L$ whereas the inverse inertia tensors $\lr$ are maximally spread in $r$.  This is an example of an \textit{uncertainty relation} associated to our duality transformations (\ref{everysec}) and (\ref{lcase}).  The factor $\delta_{jl}$ in (\ref{noninterl}) is a residual localization for $1< \r < \l$ which is induced by the $\delta_{JL}$ localization of (\ref{localL}).

Among the cyclic copy orbifolds, we note in particular the WZW cyclic orbifolds whose sectors are the \textit{orbifold affine-Sugawara} constructions$^{\rf{9}, \rf{5}}$
\namegroup{WZW}
\alpheqn
\(
(L^{ab}_{J-L})_{\gh}=\frac{\eta^{ab}}{2k+Q_{\sgb}} \delta_{JL}
\sp \gh=\oplus_{I=0}^{\l-1}\gb^I \sp \gb^I \cong \gb
\)
\(
T_{\gh}=\frac{\eta^{ab}}{2k+Q_{\sgb}} \sum_J :J_{aJ}J_{bJ}:
\)
\(
(\lr_r^{a(j)b(l)})_{\mbox{\scriptsize{\gb$_\r$}}}(\s)=\frac{1}{\r(\s)} \delta_{jl} \frac{\eta^{ab}}{2k+Q_{\sgb}} \sp \gb_{\r(\s)} = \oplus_{j=0}^{\frac{\l}{\r(\s)} -1} \gb_{\r(\s)}^j
\)
\(
\hat{T}_{{\sgb}_{\r(\s)}}=\frac{1}{\r(\s)}\frac{\eta^{ab}}{2k+Q_{\sgb}}\sum_{r=0}^{\r(\s)-1} \sum_{j=0}^{\lors-1}:\jh_{aj}^{(r)} \jh_{bj}^{(-r)}:
\)
\(
\hat{c}(\s)=c= \frac{2 \l k \ \dg}{2k+Q_{\sgb}} \sp
\hat{\Delta}_0(\s) = \frac{\l k \dg}{12 (2k+Q_{\sgb})} (1-\frac{1}{\r^2(\s)}).
\)
\reseteqn
The WZW cyclic orbifolds are the only cyclic copy orbifolds which are also $G_{\textup{\diag}(\s)}$-invariant cyclic orbifolds (see Subsec.~\ref{Gdiagsec}) and, as seen in Eq. (\ref{allsecKconj}), all orbifolds in \taz\ are paired by K-conjugation through the WZW cyclic orbifolds.

\[\]
\bu{ \textbf{Cyclic orbifolds on $\mathbf{J^{\textup{\diag}(\s)}}$}}\\
All other cyclic orbifolds (for which the stress tensor of $A(\zl)$ is not a sum of $\l$ commuting copies) are called interacting cyclic orbifolds.  We focus here on a simple class of interacting cyclic orbifolds
\(
\frac{A_{\diag}}{\z_\l} \subset \frac{A(S_\l)}{\z_\l}
\)
which we call the \textit{cyclic orbifolds on $J^{\diag (\s)}$},
\(
J^{\diag(\s)}_a (z) \equiv \sum_{j=0}^{\lors -1} \hat{J}^{(0)}_{aj} (z)
\)
because every sector of these orbifolds is constructed entirely on these currents.  The currents $J^{\diag (\s)}$ form an integral affine subalgebra$^{\rf{9}}$ in each sector $\s$ and the zero modes $Q^{\diag (\s)}$ of $J^{\diag(\s)}$ were encountered in Eq.~(\ref{gdiagJ}).

The untwisted sectors of these orbifolds are given by 
\alpheqn
\(
L_{J-L}^{ab}=L^{ab}(\l k) 
\)
\(
T=L^{ab}(\l k) : J_{a}^{\textup{\diag}}J_{b}^{\textup{\diag}}:
\)
\(
J_{a}^{\textup{\diag}} \equiv J_{a}^{\textup{\diag}(0)}=\sum_J J_{aJ}
\)
\reseteqn
where $L^{ab}(\l k)$ is any solution of the VME on simple $\gb$ at level $\l k$.

The duality transformation (\ref{mainres}) gives the stress tensors of the twisted sectors
\namegroup{iasstress}
\alpheqn
\(
\lr_r^{a(j)b(l)}(\s)=L^{ab}(\l k)\delta_{r,0} 
\)
\(
\hat{T}_\s =L^{ab}(\l k):J_a^{\textup{\diag}(\s)} J_b^{\textup{\diag}(\s)}:
\)
\(
\hat{c}(\s)=c=2\l k\eta_{ab}L^{ab}(\l k)\sp
\hat{\Delta}_0(\s)=0.
\)
\reseteqn
The cyclic orbifolds on $J^{\diag (\s)}$ are extraordinary among orbifolds (or they may not be orbifolds at all) because all their sectors have zero ground state conformal weight.  These constructions also illustrate the uncertainty relations in a different way: $L$ is maximally spread in $K=J-L$ while $\lr$ is maximally localized in $r$.

Among the cyclic orbifolds on $J^{\diag (\s)}$, one set of orbifolds
\namegroup{GJdiagoverlap}
\alpheqn
\(
T_{{\sgb}_\diag} \equiv L_{\sgb}^{ab}(\l k):J_{a}^{\textup{\diag}} J_{b}^{\textup{\diag}}: \sp L_{\sgb}^{ab}(\l k)=\frac{\eta^{ab}}{2 \l k+Q_{\sgb}}
\)
\(
\hat{T}_{{\sgb}_1, \diag (\s)}=\frac{\eta^{ab}}{2 \l k+Q_{\sgb}}:J_{a}^{\textup{\diag}(\s)} J_{b}^{\textup{\diag}(\s)}:
\)
\(
\hat{c}(\s)=c= \frac{2 \l k \dg}{2\l k +Q_{\sgb}} \sp 
\hat{\Delta}_0(\s) = 0
\)
\reseteqn
is also $G_{\diag (\s)}$-invariant (see Subsec.~\ref{Gdiagsec}).  The sectors of these orbifolds are described by a set of affine-Sugawara constructions$^{\rf{12},\rf{18},\rf{19}-\rf{21}}$ on $\{ J^{\textup{\diag}(\s)} \}$.

\[\]
\bu{ \textbf{Interacting coset orbifolds}}\\
We turn next to a simple example of the \textit{interacting coset orbifolds}
\(
\frac{(\frac{\gh}{{\sgb}_\diag})}{{\z}_{\lambda}} \subset \frac{A(S_\l)}{\z_\l} \sp \gh=\oplus_{I=0}^{\l-1} \gb^I
\sp \gb^I \cong \gb \label{cosetorb}
\)
where $\gb_\diag$ is the diagonal subalgebra of $\gh$.  Orbifolds of this type are distinct from the coset copy orbifolds
\(
\frac{\oplus_{I=0}^{\l-1}(\gb/h)^I}{\z_\l}
\)
which are special cases of the cyclic copy orbifolds in (\ref{noninter}).

The stress tensors of the interacting coset orbifolds (\ref{cosetorb}) are K-conjugate partners to the affine-Sugawara constructions on $J^{\diag (\s)}$:
\alpheqn
\(
T_{\gh/{\sgb}_\diag}=T_{\gh}-T_{{\sgb}_\diag}
\)
\(
\hat{T}_{{\sgb}_{\r(\s)}/{\sgb}_{1,\diag (\s)}}=\hat{T}_{{\sgb}_{\r (\s)}}-\hat{T}_{{\sgb}_{1,\diag (\s)}}
\)
\(
\hat{c}(\s)=c= 2 \l k \dg (\frac{1}{2k+Q_{\sgb}} - \frac{1}{2\l k +Q_{\sgb}})\sp
\hat{\Delta}_0(\s)=\frac{\l k \dg}{12\r(\s)(2k+Q_{\sgb})} (1-\frac{1}{\r^2(\s)})
\)
\reseteqn
(see (\ref{WZW}) and (\ref{GJdiagoverlap})).  The stress tensors $T$ of the untwisted sectors are the (ordinary) coset constructions$^{\rf{12},\rf{18},\rf{22}}$ $(\gh/\gb_\diag)$.  When $\r=\l$, the stress tensors $\hat{T}_\s$ are Kac-Wakimoto$^{\rf{13},\rf{14},\rf{9}}$ coset constructions, and when $\r \neq \l$ these stress tensors generalize the Kac-Wakimoto coset constructions to the case of semisimple $\gh$.  This confirms the conjecture of Ref.~\rf{9} that the Kac-Wakimoto coset constructions occur in the orbifolds (\ref{cosetorb}).  

Using the duality transformations (\ref{themap}), many other interacting coset orbifolds, for example, 
\(
\frac{(\frac{\gh}{h({\sz}_\l)})}{{\z}_{\lambda}} \sp h(\z_\l) \subset  \gh
\)
can be constructed from the coset constructions $\gh/h(\z_\l)$, where $h(\z_\l)$ is any subalgebra of $\gh$ which preserves the $\zl$-invariance of $\gh/h$.  Equivalently, one can start with the OVME constructions $\hat{T}_{{\sgb}_\l/h_\eta}$, $h_\eta\subset \gb_\l$ (see Ref.~\rf{5}) and complete these orbifolds via the duality transformations in Eq.~(\ref{sectormap}).

\[\]
\bu{ \textbf{\tad\ with prime $\l$}}\\
If $\l$ is prime then every twisted sector of any orbifold \tad\ has order $\r(\s)=\l$, and so these inverse inertia tensors take the simple form
\namegroup{primeL}
\alpheqn
\( 
\lr_r^{ab}(\s)=\frac{1}{\l} \sum_{K=0}^{\l-1} e^{-\frac{\tp N(\s)rK}{\l}} L_K^{ab}=\lr_{-r}^{ab}(\s) 
\sp \s=1,...,\l-1 \label{primex}
\)
\(
\hat{c}(\s)=c=2\l k\eta_{ab}L_0^{ab}\sp
\hat{\Delta}_0(\s)=\frac{ k\eta_{ab}}{4\l^2}(\frac{\l^2-1}{3}L_0^{ab}-\sum_{s=1}^{\l-1} csc^2(\frac{\pi Ns}{\l})\hsp{.02}L^{ab}_s)
\)
\reseteqn
where $L_K^{ab}$ is any solution of the reduced VME (\ref{DVME}).  The gauge/periodicity condition on $\lr$ in (\ref{primex}) shows that all the sectors of these orbifolds are in the OVME.

\[\]
\bu{ \tad\ \textbf{for} $\l\leq 7$}\\
Every twisted sector $\s$ of every \tad\ orbifold with $\l\leq 7$ has order $\r(\s)=2,3$ or $\l$.  As a result, we will see that, up to automorphisms of the orbifold affine algebra (see App.~C), every inverse inertia tensor of these orbifolds is in the OVME.

Sectors with order $\r(\s)=2$ have only two twist classes, $r=0$ and $1$.  Using the identity
\(
r=-r\rmod 2\sp r=0,1
\)
we see that the periodicity of $\lr$ implies the gauge/periodicity condition
\(
\lr_r^{a(j)b(l)}(\s)=\lr_{r\pm\r}^{a(j)b(l)}(\s)\rightarrow\lr_r^{a(j)b(l)}(\s)=\lr_{-r}^{a(j)b(l)}(\s)
\)
and so $\lr(\s)$ satisfies the OVME at order $\r(\s)=2$.  The sectors with order $\r(\s)=\l$ are also in the OVME (see Subsec.~\ref{OVMEsec}).

The final case, $\r(\s)=3$, is more difficult.  There are two such sectors in \tad\ with  $\l\leq 7$:
\(
\l=6\sp\s=2\textup{\ and\ }4\sp\r(2)=\r(4)=3.
\)
For brevity we will restrict our attention to the sector $\s=2$, but the same approach can be applied to $\s=4$.  The inverse inertia tensors $\lr(\s=2)$ are given by the duality transformation (\ref{mainres}):
\(
\lr_r^{a(j)b(l)}(2)=\frac{1}{3}(L_{j-l}^{ab}+e^{-\frac{\tp r}{3}}L^{ab}_{j-l+2}+e^{-\frac{4 \pi i r}{3}}L_{j-l+4}^{ab}).
\)
In particular the $j=1,\ l=0$ components can be rewritten using the reflection symmetry in (\ref{appRsym}),
\alpheqn
\(
\lr_0^{a(1)b(0)}(2)=\frac{2}{3}L_1^{ab}+\frac{1}{3}L_3^{ab}\sp
\lr_1^{a(1)b(0)}(2)=\frac{1}{3}(1+e^{-\frac{4\pi i}{3}})L_1^{ab}+\frac{1}{3}e^{-\frac{2\pi i}{3}}L_3^{ab}
\)
\(
\lr_2^{a(1)b(0)}(2)=\frac{1}{3}(1+e^{-\frac{2\pi i}{3}})L_1^{ab}+\frac{1}{3}e^{-\frac{4\pi i}{3}}L_3^{ab}
\)
\reseteqn
and we see that generically
\(
\lr_1^{a(1)b(0)}(2) \neq\lr_2^{a(1)b(0)}(2).
\)
Since $\lr$ does not satisfy the gauge condition in (\ref{appGsym}), the sector $\s=2$ is not in the OVME.

On the other hand, we can find a physically-equivalent $\lr(\s=2)\p$ that is in the OVME by using the automorphisms of the orbifold affine algebra (see App.~C) at this order.  The choice of automorphism
\(
A_{jl}(r)=\delta_{jl}e^{\frac{\tp rl}{3}}:\hsp{.4}
P_{jl}=\delta_{jl} \sp \beta(j)=j \ \textup{mod} \ 3 
\)
yields the physically-equivalent inverse inertia tensor
\(
\lr_r^{a(j)b(l)}(2)\p=\sum_{j\p,l\p}\lr_r^{a(j\p)b(l\p)}(2) A_{j\p j}(r)A_{l\p l}(-r)=\frac{1}{3}e^{\frac{\tp r(j-l)}{3}}(L_{j-l}^{ab}+e^{-\frac{\tp r}{3}}L^{ab}_{j-l+2}+e^{-\frac{4 \pi i r}{3}}L_{j-l+4}^{ab})
\)
which in fact satisfies the gauge/periodicity condition of the OVME.  

To see this, we note that there are only two cases to consider ($j=l=0$ and $j-l=\pm 1$) because $j$ and $l$ are $0$ or $1$ when $(\l/\r)=2$.  The symmetry $\lr_r=\lr_{-r}$ follows in the first case since we can use the reflection symmetry (\ref{appRsym}) to write
\(
\lr_r^{a(j)b(j)}(2)\p=\frac{1}{3}(L_{0}^{ab}+2\hsp{.03}\textup{cos}(\frac{2\pi r}{3})L^{ab}_{2})=\lr_{-r}^{a(j)b(j)}(2).
\)
For the case $j-l=\pm 1$, we can again use (\ref{appRsym}) to obtain
\(
\lr_r^{a(j)b(j\pm 1)}(2)\p=\frac{1}{3}(2\hsp{.03}\textup{cos}(\frac{2\pi r}{3})L_{1}^{ab}+L_3^{ab})=\lr_{-r}^{a(j)b(j\pm 1)}(2)
\)   
so $\lr(2)\p$ is in the OVME.  

This completes the demonstration that, up to automorphisms of the orbifold affine algebra, all sectors of all orbifolds \tad \ with $\l\leq7$ are in the OVME.  When $\l\geq 8$ however, many twisted sectors of \tad \ can only be found in the cyclic OVME (\ref{COVMEc}).


\end{document}